\documentclass[12pt,a4paper]{article}

\usepackage{epsfig}
\usepackage{exscale}
\def\lsim{\raise0.3ex\hbox{$<$\kern-0.75em\raise-1.1ex\hbox{$\sim$}}}
\def\gsim{\raise0.3ex\hbox{$>$\kern-0.75em\raise-1.1ex\hbox{$\sim$}}}
\newcommand{\<}{\langle}
\renewcommand{\>}{\rangle}
\newcommand{\be}{\begin{equation}}
\newcommand{\ee}{\end{equation}}
\newcommand{\ba}{\begin{eqnarray}}
\newcommand{\ea}{\end{eqnarray}}

\def\spose#1{\hbox to 0pt{#1\hss}}
\def\ltapprox{\mathrel{\spose{\lower 3pt\hbox{$\mathchar"218$}}
 \raise 2.0pt\hbox{$\mathchar"13C$}}}
\def\gtapprox{\mathrel{\spose{\lower 3pt\hbox{$\mathchar"218$}}
 \raise 2.0pt\hbox{$\mathchar"13E$}}}
\def\phv{\vec \phi}

\def\NT{N_\tau}
\def\nt{\ifmmode\NT\else$\NT$\fi}
\def\NS{N_\sigma}
\def\ns{\ifmmode\NS\else$\NS$\fi}

\def\PRep{{ Phys.\ Rep.\ }}

\def\p{^\prime}
\def\v{\vec}

\def\n{\noindent}

\begin{document}
\begin{titlepage} 
\thispagestyle{empty}

 \mbox{} \hfill BI-TP 2009/26\\
 \mbox{} \hfill November 2009\\ %\today 
% \mbox{} \hfill cond-mat/0209492v2
\begin{center}
\vspace*{0.8cm}
{{\Large \bf Longitudinal and transverse spectral functions\\
 in the three-dimensional $O(4)$ model\\}}\vspace*{1.0cm}
{\large J. Engels and O. Vogt}\\ \vspace*{0.8cm}
\centerline {{\em Fakult\"at f\"ur Physik, 
    Universit\"at Bielefeld, D-33615 Bielefeld, Germany}} \vspace*{0.4cm}
\protect\date \\ \vspace*{0.9cm}
{\bf   Abstract   \\ } \end{center} \indent
We have performed a high statistics simulation of the $O(4)$ model
on a three-dimensional lattice of linear extension $L=120$ for small 
external fields $H$. Using the maximum entropy method we analyze
the longitudinal and transverse plane spin correlation functions 
for $T<T_c$ and $T\ge T_c$. In the transverse case we find for all $T$ 
and $H$ a {\em single} sharp peak in the spectral function, whose 
position defines the transverse mass $m_T$, the correlator is that of a
free particle with mass $m_T$. In the longitudinal case we find in the
very high temperature region also a single sharp peak in the spectrum. 
On approaching the critical point from above the peak broadens somewhat
and at $T_c$ its position $m_L$ is at $2m_T$ for all our $H-$values. 
Below $T_c$ we find still a significant peak at $\omega=2m_T$ and at 
higher $\omega-$values a continuum of states with several smaller peaks
with decreasing heights. This finding is in accord with a relation of
Patashinskii and Pokrovskii between the longitudinal and the transverse
correlation functions. We test this relation and its range of 
applicability in the following. As a by-product we calculate critical
exponents and amplitudes and confirm our former results.    

\vfill \begin{flushleft} 
Spectral function\\
\noindent{\rule[-.3cm]{5cm}{.02cm}} \\
\vspace*{0.2cm}
Corresponding author: J\"urgen Engels\\
Fakult\"at f\"ur Physik, Universit\"at Bielefeld, D-33615 Bielefeld, Germany\\ 
Tel.: +49 521 1065317, Fax: +49 521 1062961\\
E-mail addresses: engels@physik.uni-bielefeld.de, 
o-vogt@t-online.de
\end{flushleft} 
\end{titlepage}
%\end{document}

%%%%%%%%%%%%%%%%%%%%%%%%%%%%%%%%%%%%%%%%%%%%%%%%%%%%%%%%%%%%%%%%%%%%%%%%%%%%%%%%

\section{Introduction}

%%%%%%%%%%%%%%%%%%%%%%%%%%%%%%%%%%%%%%%%%%%%%%%%%%%%%%%%%%%%%%%%%%%%%%%%%%%%
$O(N)$ spin models with $N>1$ possess two types of two-point correlation 
functions, one for the transverse and and one for the longitudinal spin
component, defined relative to the direction of the external field $\vec H$.
Correspondingly, there exist also two distinct susceptibilities and 
correlation lengths. The difference and interplay between the two sets of
observables is of interest not only for the critical behaviour near to 
$T_c$, the critical temperature, but as well for the study of the 
singularities induced by the existence of the massless Goldstone modes in
these $O(N)$ spin models for dimension $2<d\le 4$ and all
$T<T_c$ \cite{Zinn-Justin:1996,Anishetty:1995kj}. 

In a previous paper \cite{Engels:2003nq} we studied the correlation
lengths which govern the exponential decay of the two correlation functions
in the three-dimensional $O(4)$-model on lattices of linear extensions
$L=48-120$. We were able to confirm the predicted singular
Goldstone behaviour \cite{Fisher:1985,Patashinskii:1973} of the transverse 
correlation length near the coexistence line, $T<T_c,\, H=0$ and to 
determine the scaling function of $\xi_T$. For the longitudinal correlation
length the situation was different. In the high temperature region the 
scaling function could be calculated, in the low temperature phase,
$T<T_c\,$, we were however unable to reliably estimate $\xi_L$, the data
were not even scaling. This was ascribed to a spectrum of higher states
which contribute to the longitudinal correlators below $T_c$. In this paper
we want to calculate the spectral functions of the two correlators and to
actually find these states above the ground states $m_{T,L}=1/\xi_{T,L}$.
Here, not only the spectrum of the longitudinal correlator is of interest,
but as well the transverse spectrum. Spin-wave theory assumes, that 
long-wavelength transverse fluctuations dominate for small fields in thermal
equilibrium below $T_c$ and that these fluctuations are describable by the
Gaussian model or free field functional. This assumption can be tested
readily by comparing the transverse correlator to the known Gaussian form.
Of course, no higher state should then appear in the spectrum. 

The resulting dependence on $H$ of the lowest states $m_T$ and $m_L$ 
at fixed $T$ can be utilized to test critcal behaviour and the effects
of massless Goldstone modes. At the critical point we had already found
in Ref.\ \cite{Engels:2003nq} that $m_L = 2m_T$, like for $O(2)$
\cite{Cucchieri:2002hu}. The equation between the two masses or 
correlation lengths follows from a relation between the transverse and
longitudinal correlation functions \cite{Patashinskii:1973,Fisher:1985}
which should be valid for small $H$ in the whole low temperature phase.
We shall discuss and test this relation explicitly for its range of 
applicability. In order to achieve these goals, we simulate the 
$O(4)$-invariant nonlinear $\sigma$-model with high statistics on a
lattice with linear extension $L=120$. We have chosen this specific 
model for several reasons: first because we want to clarify the open
points raised in Ref.\ \cite{Engels:2003nq}, second
because in contrast to the corresponding $O(2)$-model corrections to 
scaling are negligible here, and third because the model is of relevance 
for quantum chromodynamics (QCD) with two degenerate light-quark flavours
at finite temperature, since its phase transition is supposed
to belong to the same universality class as the chiral transition of QCD
\cite{Engels:2005rr,Basile:2005hw,Ejiri:2009ac}.

The $O(N)$-model which we study here is defined by the Hamiltonian
\be
\beta\,{\cal H}\;=\;-J \,\sum_{<{\v x},{\v y}>}\phv(\v x)\cdot
\phv(\v y) \;-\; {\v H}\,\sum_{{\v x}} \phv(\v x) \;,
\ee
where ${\v x}$ and ${\v y}$ are nearest-neighbour sites on a 
three-dimensional hypercubic lattice with periodic boundary conditions,
and $\phv(\v x)$ is a $N$-component spin vector at site ${\v x}$ with
length 1. It is convenient to give the following formulas still for
general $N$, though we use in our simulations $N=4$. The magnetization 
vector $\v M$ is the expectation value of the lattice average $\phv$ of
the local spins 
\be
\v M = \< \phv \>~, \quad {\rm with}\quad  \phv = 
{1 \over V} \sum_{{\v x}} \phv(\v x)~.
\ee
Here, $V=L^3$ and $L$ is the number of lattice points per direction,
the lattice spacing $a$ is fixed to 1.
Due to the invariance of ${\cal H}_0$, the $\v H$-independent part
of ${\cal H}$, under $O(N)$-rotations of $\phv(\v x)$, the magnetization
vector aligns with $\v H$
\be
\v M = M \v n ~,\quad {\rm with}\quad \v n=\v H/ H~.
\ee
It is therefore appropriate also to decompose the spin vectors $\phv(\v x)$
into longitudinal (parallel to the magnetic field $\v H$) and transverse
components 
\be
\phv(\v x)\; =\; \phi^{\parallel}(\v x)~ \v n +
\phv^{\perp}(\v x) ~.
\ee
The order parameter of the system, the magnetization $M$, is then
the expectation value of the lattice average $\phi^{\parallel}$
of the longitudinal spin components 
\be
M =\;\<\, \phi^{\parallel} \,\>~, \quad {\rm and}\quad \v M^{\perp}
=\;\<\, \phv^{\perp} \,\>=\;0~.
\ee
Owing to the $N$ components of $\v M$ and $\v H$ one is led to a
whole matrix of susceptibilities
\be
\chi_{ab} = \frac{\partial M_a}{\partial H_b} 
= \frac{\partial  \< \phi_a \>}{\partial H_b} = V\bigm(
\< \phi_a \phi_b \> - \< \phi_a \>\< \phi_b \> \bigm)~,
\label{cdef}
\ee
with $a,b=1,\dots,N$. The matrix can be expressed with only two 
quantities
\be
\chi_{ab} = \chi_L n_a n_b + \chi_T (\delta_{ab} -n_a n_b)~,
\label{cmat}
\ee
the longitudinal and the transverse susceptibility. The total
susceptibility is
\be
\chi_{tot} = Tr (\chi) = \chi_L + \chi_T (N-1) 
           = V\bigm(\< \phv^2 \> - \< \phv \>^2 \bigm)~.
\label{ctot}
\ee
From the last equation and 
\be
\chi_L = \v n \chi \v n = V\bigm(\<\, \phi^{\parallel2} \,\>-M^2\bigm)~,
\label{chil}
\ee
one finds, as expected, that the transverse susceptibility corresponds
to the fluctuation per component of the lattice average $\phv^{\perp}$
of the transverse spin components
\be
\chi_T =  {V \over N-1} \<\, \phv^{\perp2} \,\> ~.
\label{chit}
\ee
The two susceptibilities can as well be obtained from the equations
\be
\chi_L = \frac{\partial M}{\partial H}~,  \quad {\rm and}\quad \chi_T
=\frac{M}{ H}~. \label{cmoh}
\ee

The remainder of the paper is organized as follows. We start with a 
detailed discussion of the correlation functions and their properties,
the formula connecting the two types of correlators and their spectral
representations. In Section 3 we address the critical behaviour of the
observables. In the following
Section 4 we describe some details of our simulations and the used 
maximum entropy method (MEM). Then we analyze our data in the different
$T$-regions: below $T_c$, at $T_c$ and in the high temperature region;
first separately for the transverse and longitudinal correlation functions,
subsequently we investigate their interplay. We close with a summary and
the conclusions. 

%%%%%%%%%%%%%%%%%%%%%%%%%%%%%%%%%%%%%%%%%%%%%%%%%%%%%%%%%%%%%%%%%%%%%%%%%%%%%%%%

\section{The correlation functions}
\label{section:Corfu}

%%%%%%%%%%%%%%%%%%%%%%%%%%%%%%%%%%%%%%%%%%%%%%%%%%%%%%%%%%%%%%%%%%%%%%%%%%%%%%%%

\subsection{2-Point correlators}
\label{section:2Pocor}

%%%%%%%%%%%%%%%%%%%%%%%%%%%%%%%%%%%%%%%%%%%%%%%%%%%%%%%%%%%%%%%%%%%%%%%%%%%%%%%%
As for the susceptibilities, there exists a matrix of 2-point connected
correlation functions 
\be
G_{ab}(\v x_1, \v x_2) = \< \phi_a(\v x_1)\phi_b(\v x_2)\>_c
 = \< \phi_a(\v x_1)\phi_b(\v x_2)\> -
\< \phi_a(\v x_1)\> \<\phi_b(\v x_2)\>~.
\label{gdef}
\ee 
Furthermore, because of translation invariance we have 
\be
\< \phv(\v x) \> =  \< \phv\>~,\quad {\rm and} \quad G_{ab}(\v x_1, \v x_2)
= G_{ab}(\v x_2- \v x_1)~. \label{transl}
\ee
The correlation functions are related to the susceptibilities by
\be
\chi_{ab} = \sum_{{\v x}} G_{ab} (\v x)~.
\label{fludiss}
\ee
The matrix $G$ must have the same form as the matrix $\chi$
\be
 G_{ab} = G_L n_a n_b+ G_T (\delta_{ab} -n_a n_b)~,
\label{gmat}
\ee
leading to
\ba
G_L (\v x) &=& \< \phi^{\parallel}(0) \phi^{\parallel}(\v x) \>
- \< \phi^{\parallel} \>^2~, \label{gpl}\\
G_T (\v x) &=& \frac{1}{N-1}\< \phv^{\perp}(0)\phv^{\perp}(\v x)\>~.
\label{gpt}
\ea
%The total correlation function is
%\be
%G_{tot}(\v x)= G_L(\v x) +(N-1)G_T(\v x)=  \< \phv(0) \phv(\v x) \>
%- \< \phv \>^2~. \label{gptot}
%\ee
As mentioned already, the large distance behaviour of the correlation
functions $G_{L,T}$ is determined by the respective exponential correlation 
lenghts $\xi_{L,T}$
\be
G_{L,T}({\v x}) \sim \exp (-|\v x|/\xi_{L,T})~.
\label{xiexp}
\ee
This is true for all temperatures $T$ (the coupling $J$ acts here as inverse
temperature, that is $J=1/T$) and external fields $H$, except on the
coexistence line $H=0, T<T_c$ and at the critical point, where the 
correlation lengths diverge.
%%%%%%%%%%%%%%%%%%%%%%%%%%%%%%%%%%%%%%%%%%%%%%%%%%%%%%%%%%%%%%%%%%%%%%%%%%%%%%%%

\subsection{2-Plane correlators}
\label{section:2Plcor}

%%%%%%%%%%%%%%%%%%%%%%%%%%%%%%%%%%%%%%%%%%%%%%%%%%%%%%%%%%%%%%%%%%%%%%%%%%%%%%%%
Instead of the 2-point correlation functions we shall actually measure
2-plane correlation functions. They are defined as the connected correlation
functions of spin averages over planes. For example, the spin average over
the $(x,y)$-plane at position $z$ is 
\be
{\v S}(z) =  { 1 \over L^2 } \sum_{x,y} \phv(\v x)~.
\label{spinav}
\ee
The general 2-plane connected correlation functions are then
\be
D_{ab}(z)= \sum_{x,y} G_{ab}(\v x) = L^2 \<S_a(0)S_b(z)\>_c~.
\label{ddef}
\ee
Also the expectation values of the longitudinal and transverse
components of ${\v S}(z)$ are independent of $z$ and equal to those 
of the corresponding lattice averages 
\be
\langle S^{\parallel}(z) \rangle =\langle \phi^{\parallel} \rangle =M~,
\quad {\rm and} \quad \langle  {\v S}^{\perp}(z)\rangle =
 \langle\phv^{\perp}\rangle =0~.
\label{sexp}
\ee 
The respective longitudinal and transverse plane-correlation 
functions $D_{L,T}(z)$ are
\ba
D_L(z)\!\!& =&\!\! L^2 \bigm(\langle S^{\parallel}(0)
S^{\parallel}(z) \rangle - M^2 \bigm)~,\label{planel}\\
D_T(z)\!\!& =&\!\! {L^2 \over N-1}
\langle {\v S}^{\perp}(0) {\v S}^{\perp}(z)\rangle ~.
\label{planet}
\ea
Here, $z$ is the distance between the two planes. Instead of choosing 
the $z$-direction as normal to the plane one can as well take the 
$x$- or $y$-directions. Accordingly, we enhance the accuracy of the 
correlation function data by averaging over all three directions 
and all possible translations. The correlators are symmetric
and periodic functions of the distance $\tau$ between the planes,
the factor $L^2$ on the right-hand sides of (\ref{planel})
and (\ref{planet}) ensures the relation
\be
\chi_{L,T} = \sum_{\tau=0}^{L-1} D_{L,T}(\tau)~.
\label{fluc}
\ee
Like the point-correlation functions in Eq.\ (\ref{xiexp}) the 
plane correlators $D_{L,T}(\tau)$ decay exponentially.  
%%%%%%%%%%%%%%%%%%%%%%%%%%%%%%%%%%%%%%%%%%%%%%%%%%%%%%%%%%%%%%%%%%%%%%%%%%%%%%%%

\subsection{The Gaussian model}
\label{section:Gauss}

%%%%%%%%%%%%%%%%%%%%%%%%%%%%%%%%%%%%%%%%%%%%%%%%%%%%%%%%%%%%%%%%%%%%%%%%%%%%%%%%
We noted already in the introduction, that the Gaussian model is assumed
to describe the long-wavelength transverse fluctuations. Therefore, we 
shortly discuss, following Ref.\ \cite{Patashinskii:1979}, the aspects of
the model which are relevant for our
work, first in $d$ dimensions, finite volume and the continuum limit.
Here, $\phi(\v x)$ is a single component, real scalar field with the
Hamiltonian
\be
\beta {\cal H}_G = \int d^d x \bigm[ \frac{1}{2}c_s (\nabla \phi)^2
+\frac{1}{2}a_s \phi^2 \bigm]~,
\label{freef}
\ee 
where $c_s$ is the so-called {\it stiffness} and $a_s$ is proportional 
to the square of the mass of the field. Let us introduce the
Fourier transform $\tilde\phi(\v k)$ of the field
\ba
\tilde\phi(\v k) &\!\! =\!\! & \int d^d x e^{-i\v k \v x} \phi(\v x)~,
\label{phik} \\
\phi(\v x) &\!\! =\!\! & \frac{1}{V} \sum_{\v k} e^{i\v k \v x}
\tilde\phi(\v k)~. \label{phix}
\ea
For periodic boundary conditions, the components of $\v k$ are
restricted to $k_i= 2\pi n_i/L$ with $n_i=0,\pm 1,\dots,$ and because
$\phi$ is real we have $\tilde\phi(\v k)^{\star}=\tilde\phi(-\v k)$.
In terms of $\tilde\phi(\v k)$ the Hamiltonian is
\be
\beta {\cal H}_G =  \frac{1}{2V} \sum_{\v k} |\tilde\phi(\v k)|^2 
(c_s {\v k}^2 +a_s)~, \label{H_G}
\ee
that is the $\tilde\phi(\v k)$ decouple, 
\be
\<\tilde\phi(\v k)\> =0~, \quad {\rm and} \quad \<|\tilde\phi(\v k)|^2\>
= {V \over c_s {\v k}^2 +a_s}~. \label{basic1}
\ee
That leads to the results
\be
\<\phi(\v x)\>=0~, \quad {\rm and} \quad \<\phi(0)\phi(\v x)\> =
 \frac{1}{V} \sum_{\v k} {e^{i\v k \v x} \over c_s {\v k}^2 +a_s}~.
\label{basic2}
\ee
The latter expectation value is thus directly the 2-point connected
correlator of $\phi(\v x)$.
It is easy to evaluate the Gaussian correlator for $d=1$, the case of 
the 2-plane correlation function
\be
D(\tau) = \frac{1}{L} \sum_{n} {e^{ik_n \tau} \over c_s k_n^2 +a_s}
= \frac{1}{Lc_s} \sum_{n} {e^{ik_n \tau} \over  k_n^2 + m^2}~.
\label{Dtau}
\ee
Here, $k_n=2\pi n/ L$ and the mass is $m=(a_s/c_s)^{1/2}$. Using the
well-known relation
\be
\frac{1}{L}\sum_{n} {e^{ik_n \tau} \over x +ik_n} =
{e^{-x\tau} \over 1 -e^{-Lx}}~,
\label{period}
\ee
for $x=\pm m$, we obtain
\be
D(\tau) = \frac{1}{2mc_s} \cdot {e^{-m\tau}+e^{-m(L-\tau)} \over 1-e^{-mL}}
~. \label{Dm}
\ee
In addition, we can identify $a_s$ with the inverse of the susceptibility
\be
\chi = \int\limits_0^L D(\tau)d\tau = \frac{1}{m^2c_s} = \frac{1}{a_s}~.
\label{chia}
\ee
In the infinite volume limit, $V\to \infty$, the Gaussian correlator
is
\be
G(\v x) = \frac{1}{c_s}\int\limits_{-\infty}^{\infty}
\frac{d^d k}{(2\pi)^d}~{e^{i\v k \v x} \over {\v k}^2 + m^2}~,\quad
{\rm and} \quad {\tilde G}(\v k)=\frac{1}{c_s}~
{1 \over {\v k}^2 + m^2}~. \label{gandgf}
\ee
The integral is known for general $d~$
\cite{Patashinskii:1979,Binney:1992}
\be
G(\v x) = \frac{1}{(2\pi)^{d/2}c_s}\left( {m \over |\v x|}\right)^{d/2-1}
K_{d/2-1} (m|\v x|)~,  \label{general}
\ee 
where $K_{\nu}(z)$ is a modified Bessel function. We are interested here
in the two cases
\ba
G(\v x) &\!\!=\!\!& \frac{e^{-m|\v x|}}{4\pi c_s |\v x| }~,\qquad d=3~,
\label{dim3}\\
D(\tau) &\!\!=\!\!& \frac{e^{-m\tau}} {2mc_s} ~, \quad \qquad d=1~.
\label{dim1}
\ea
The last equation can also be obtained from Eq.\ (\ref{Dm}) for 
$L\to \infty$. 
%%%%%%%%%%%%%%%%%%%%%%%%%%%%%%%%%%%%%%%%%%%%%%%%%%%%%%%%%%%%%%%%%%%%%%%%%%%%%%%%

\subsection{The relation between the transverse and longitudinal \\
correlation functions}
\label{section:tlrel}

%%%%%%%%%%%%%%%%%%%%%%%%%%%%%%%%%%%%%%%%%%%%%%%%%%%%%%%%%%%%%%%%%%%%%%%%%%%%%%%%
The connection between the two correlation functions was first 
established in a paper by Patashinskii and Pokrovskii
\cite{Patashinskii:1973}. The main arguments leading to this relation are
the following. The transverse susceptibility $\chi_T$, Eq.\ (\ref{cmoh}),
obviously becomes infinite for all $T<T_c$ and $H\to 0$, due to the 
non-zero spontaneous magnetization $M$ in the low temperature phase. The
instability comes about because of infinitesimal long-wavelength transverse
fluctuations which change the direction of $\phv$ but not its length $\phi$.
For a fluctuation 
\be
\delta \phv = \phv - \<\phv \>~, \quad \delta \phi^{\parallel} =
\phi^{\parallel} -M~, \quad \delta \phv^{\perp} = \phv^{\perp}~,
\label{delp}
\ee 
the length change is
\be
\delta \phv^{\;2}= \delta \phi^2 = \phi^2 - M^2 = (\delta \phi^{\parallel})^2
+ (\delta \phv^{\perp})^2 + 2M \delta \phi^{\parallel}~.
\label{lech}
\ee
To lowest order in $\delta \phi^{\parallel}$ length conserving   
fluctuations 
\be
\delta \phv^{\;2}\approx (\delta \phv^{\perp})^2 + 2M \delta \phi^{\parallel}
=0~, \label{lecon}
\ee
entail longitudinal fluctuations of size
\be
\delta \phi^{\parallel}= - {(\delta \phv^{\perp})^2 \over 2M }~.
\label{lonflu}
\ee
Patashinskii and Prokrovskii extend this "principle of conservation
of the modulus" also to local fluctuations, that is to non-uniform
spin-waves with long wavelengths
\be
\delta \phi^{\parallel}(\v x) = - {\bigm(\delta \phv^{\perp}(\v x)
\bigm)^2 \over 2M }~. \label{localfluc}
\ee
The longitudinal correlation function $G_L(\v x)$ can then be
expressed as
\ba
G_L(\v x)&\!\! =\!\!& \< \phi^{\parallel}(0) \phi^{\parallel} (\v x) \>_c
=\< \delta \phi^{\parallel}(0)\delta \phi^{\parallel}
(\v x) \> \\
&\!\! =\!\!& \frac{1}{4M^2} \left< \bigm(\delta \phv^{\perp}(0) \bigm)^2
 \bigm(\delta \phv^{\perp}(\v x)\bigm)^2 \right>_c~.
\label{glfluc}
\ea
In the last expectation value we have explicitly noted that the
connected value has to be taken. As already mentioned, the transverse
fluctuations $\delta \phv^{\perp}(\v x)=\phv^{\perp}(\v x)$ are assumed
to have the Gaussian form of Eq.\ (\ref{freef})
\be
\beta {\cal H}_T = \int d^3 x \bigm[ \frac{1}{2}c_s (\nabla 
\phv^{\perp})^2
+\frac{1}{2\chi_T} \phv^{\perp 2} \bigm]~.
\label{hyddes}
\ee
The stiffness $c_s$ depends on the temperature and is to lowest order
independent of $H$. It measures the resistance of the system against 
non-uniform rotations of $\phv(\v x)$. In the disordered phase, where the 
correlation length is finite, distant points can have different
orientations and $c_s$ should be essentially 1. From Eq. (\ref{hyddes})
it is clear that the components of $\phv^{\perp}$ do not interact,
since 
\be
(\nabla\phv^{\perp})^2 = \sum\limits_{i=1}^3\sum\limits_{a=1}^N
\left( {\partial \phi^{\perp}_a \over \partial x_i} \right)^2
\label{nabla}
\ee 
and the components $\phi^{\perp}_a$ decouple.
We may therefore use Wick's theorem to reduce the expectation value
in Eq.\ (\ref{glfluc})
\be
\< \phi^{\perp 2}_a (0) \phi^{\perp 2}_b (\v x) \>
= 2 \< \phi^{\perp}_a (0) \phi^{\perp}_b (\v x) \>^2 +
\< \phi^{\perp 2}_a (0)\>\<\phi^{\perp 2}_b (\v x)\> ~,
\label{wick}
\ee
so that
\be
G_L(\v x) = \frac{1}{2M^2} \sum\limits_{a,b=1}^N 
\< \phi^{\perp}_a (0) \phi^{\perp}_b (\v x) \>^2= 
\frac{1}{2M^2} \sum\limits_{a,b=1}^N \bigm(
 G_T (\v x) (\delta_{ab}-n_a n_b) \bigm)^2~,
\label{nearly}
\ee
where we have used Eq.\ (\ref{gmat}). That finally leads to 
\be
G_L (\v x) = \frac{N-1}{2M^2}\; G_T^2(\v x)~.
\label{PP}
\ee
Let us briefly discuss the consequences of Eq.\ (\ref{PP}), 
the PP-relation, in the infinite volume limit. 
The transverse correlation function is then given by Eq.\ (\ref{dim3})
%\be
%G_T(\v x) = \frac{e^{-m_T|\v x|}}{4\pi c_s |\v x| }~,
%\label{gtsingle}
%\ee
with the mass $m_T$ and 
\be
 m_T^2 = \frac{1}{\chi_T c_s} = \frac{H}{Mc_s}~,
\label{mt}
\ee
whereas the longitudinal correlation function is
\be
G_L(\v x) = \frac{N-1}{2M^2}\;\frac{e^{-2m_T|\v x|}}
{(4\pi c_s |\v x|)^2}~.
\label{glPP}
\ee
The Fourier transform of $G_L$ is
\be
\tilde G_L(\v k) =  \frac{N-1}{2M^2}\;\frac{1}{4\pi c_s^2 k} 
\arctan \frac{k}{2 m_T}~, \label{gltilde} 
\ee
where $k=|\v k|$. It is instructive to compute the corresponding
longitudinal susceptibility
\be
\chi_L = \tilde G_L(\v k=0) = \frac{N-1}{2M^2}\;\frac{1}{4\pi c_s^2 2m_T}
~, \label{cl1}
\ee
or
\be
\chi_L = \frac{N-1}{16\pi (Mc_s)^{3/2}} \cdot H^{-1/2}~.
\label{chigold}
\ee
The last equation implies the divergence of $\chi_L\sim H^{-1/2}$ for
$H\to 0$ and $T<T_c$. That is exactly the singular Goldstone
behaviour predicted in Refs.\ \cite{Wallace:1975} and
\cite{Zinn-Justin:1996,Anishetty:1995kj} from field theory. Furthermore,
we find from Eq.\ (\ref{mt}) that near the coexistence line
\be
m_T \sim H^{1/2}~, \quad {\rm or} \quad \xi_T \sim H^{-1/2}~,
\label{xigold}
\ee
and from Eq.\ (\ref{PP}), if $G_T(\v x)$ is decaying exponentially
\be
\xi_T=2\, \xi_L~.
\label{exdec}
\ee

In order to test all these relations with our data we need 
the lattice versions of the one-pole correlation functions, Eqs.\
(\ref{dim3}) and (\ref{dim1}), and the corresponding predictions from
the PP-relation (\ref{PP}). On a lattice with periodic boundary 
conditions the Fourier transforms are
\be
G(\v x) = \frac{1}{L^3} \sum_{\v k} e^{i\v k \v x}\,{\tilde G}(\v k)~,
\quad {\tilde G}(\v k) = \sum_{\v x} e^{-i\v k \v x}\,G(\v x)~,
\label{latfou}
\ee
where $\v x=(n_1,n_2,n_3)$ with $n_i=0,\dots,L-1$ and $\v k = 2\pi
(m_1,m_2,m_3)/L$ with $m_i= 0,\pm 1,\dots, \pm (L/2-1), L/2$ for even $L\,$.
Evidently, the 2-plane correlation function $D_T$ is proportional to the 
sum of exponentials
\be
D_T(\tau) \sim \bigm( e^{-m_T\tau}+e^{-m_T(L-\tau)} \bigm)~.
\label{dprop} 
\ee 
The normalization factor is determined by Eq.\ (\ref{fluc}), so that
\be
D_T(\tau) =\chi_T \,\tanh \left(\frac{m_T}{2}\right) \cdot
 { e^{-m_T\tau}+e^{-m_T(L-\tau)} \over 1 - e^{-m_TL}}~.
\label{dlat}
\ee
For small $m_T$ the prefactor
\be
\kappa = \chi_T \tanh \left(\frac{m_T}{2}\right) 
\label{kap}
\ee  
coincides with that of Eq.\ (\ref{Dm}). The Fourier transform of $D_T$ is
\be
\tilde D_T (k) = \kappa \cdot {2 \sinh (m_T) \over 4 \sinh^2\left(
\frac{m_T}{2} \right) + 4 \sin^2 \left( \frac{k}{2} \right) }~.
\label{dflat}
\ee
As in Ref.\ \cite{Engels:1994ek} we extend the form (\ref{dflat}) to the
three-dimensional Fourier transform
\be
\tilde G_T (\v k) = \kappa \cdot {2 \sinh (m_T) \over 4 \sinh^2\left(
\frac{m_T}{2} \right) + 4 \sum\limits_{i=1}^3\sin^2 \left( \frac{k_i}{2}
\right) }~,
\label{gflat}
\ee
from which $G_T(\v x)$ is obtained, using Eq.\ (\ref{latfou}). Once we have
determined $m_T$ from MEM or a direct fit of the transverse 2-plane
correlator data to Eq.\ (\ref{dlat}), we can compare the longitudinal
data to the following predictions from the PP-relation. Let us first
compute the longitudinal susceptibility
\be
\chi_L  = \frac{N-1}{2M^2}\; \sum_{\v x}G_T^2(\v x)~. 
%  &\!\! =\!\!&  \frac{N-1}{2M^2}\;\frac{1}{L^3} {\sum_{\v k}}\p\tilde  G_T^2 (\v k)
%  &\!\! =\!\!& 
\label{cpp1}
\ee
From
\be
 \sum_{\v x}G_T^2(\v x) = \frac{1}{L^3} {\sum_{\v k}} {\sum_{\v k\p}}
\delta_{\v k+\v k\p,\,0}\; \tilde G_T (\v k)\tilde G_T (\v k\p)
=\frac{1}{L^3} {\sum_{\v k}}\p \tilde  G_T (\v k)\tilde  G_T (-\v k)~,
\ee
where the prime on the sum indicates that there is no contribution from
$k_i=\pi\,$, we find
\ba
\chi_L &\!\! =\!\!&  \frac{N-1}{2M^2}\;\frac{1}{L^3} {\sum_{\v k}}\p
\tilde  G_T^2 (\v k)\label{cpp2} \\
 &\!\! =\!\!& \frac{N-1}{2H^2L^3}\;\sinh^4 \left(\frac{m_T}{2}\right)
 {\sum_{\v k}}\p \left[\sinh^2 \left(\frac{m_T}{2}\right)
+   \sum\limits_{i=1}^3\sin^2 \left( \frac{k_i}{2} \right) \right] ^{-2}~.
\label{cpp3}
\ea
In the last equation we have used Eq.\ (\ref{kap}) and $\chi_T=M/H$.
Similarly, we can find the longitudinal 2-plane correlator from
\be
D_L(z) = \sum_{x,y} G_L(\v x) =  \frac{N-1}{2M^2}\;\sum_{x,y}
 G_T^2(\v x)~. \label{dlz}
\ee
Here, we have
\be
D_L(z) =  \frac{N-1}{2M^2L^4}\; \sum_{k_z,k_z\p}
e^{i(k_z+k_z\p )z}{\sum_{k_x,k_y}}\p \tilde G_T(k_x,k_y,k_z)
\tilde G_T(k_x,k_y,k_z\p)~, 
\label{dpp1}
\ee
or
\ba
D_L(\tau)&\!\!\!\!\! =\!\!\!\!\!&  \frac{N-1}{2H^2L^4}\;\sinh^4 
\left(\frac{m_T}{2}\right)
 \sum_{k_3,k_3\p}e^{i(k_3+k_3\p )\tau} {\sum_{k_1,k_2}}\p
\left[\sinh^2 \left(\frac{m_T}{2}\right)
+   \sum\limits_{i=1}^3\sin^2 \left( \frac{k_i}{2} \right) \right]^{-1}
\nonumber \\
&\!\!\!\!\!\!\!\! & \times\left[\sinh^2 \left(\frac{m_T}{2}\right)
+   \sum\limits_{i=1}^2\sin^2 \left( \frac{k_i}{2} \right)
+ \sin^2 \left( \frac{k_3\p}{2} \right)\right]^{-1} ~. \label{dpp2}
\ea
%%%%%%%%%%%%%%%%%%%%%%%%%%%%%%%%%%%%%%%%%%%%%%%%%%%%%%%%%%%%%%%%%%%%%%%%%%%%%%%%

\subsection{The spectral functions}
\label{section:specfu}

%%%%%%%%%%%%%%%%%%%%%%%%%%%%%%%%%%%%%%%%%%%%%%%%%%%%%%%%%%%%%%%%%%%%%%%%%%%%%%%%
We define the spectral function $A(\omega)$ of the 2-plane correlation
function by the integral equation 
\be
D(\tau) = \int\limits_0^{\infty}d\omega A(\omega) K(\omega,\tau)~,
\label{specfun}
\ee 
where the integral kernel is
\be
K(\omega,\tau)={e^{-\omega\tau}+e^{-\omega(L-\tau)} \over 1-e^{-\omega L}}
 = {\cosh(\omega(\tau-L/2)) \over \sinh (\omega L/2)}~. 
\label{kernel0}
\ee
This choice of the kernel is motivated on one hand by the lattice form
of the 2-plane correlators as obtained in the transfer matrix (TM)
formalism \cite{Engels:1994ek} and second from the appearance of the 
same factor $K(\omega,\tau)$ in the continuum, Eq.\ (\ref{Dm}),
and lattice forms, Eq.\ (\ref{dlat}), of the one-pole correlation function.
Moreover, this kernel corresponds to the one known from finite temperature
QCD \cite{Asakawa:2000tr,Datta:2003ww}, where $L$ is replaced by 
$\beta=1/T$. Yet, we are free to redefine the kernel and the spectral
function by a convenient factor $q(\omega)$ and its inverse such
that $D(\tau)$ remains invariant. In a paper by Aarts et al.\ 
\cite{Aarts:2007wj} this freedom has been used to remove numerical 
instabilities in Bryan's MEM-algorithm \cite{Bryan:1990} at small
$\omega$ which are due to the singularity of the kernel at $\omega=0$  
by changing the kernel with the factor (in our notation)
\be
q(\omega)= \omega L/2~, \quad \bar K =q K ~,\quad \bar A =A/q~.
\label{aarts}
\ee
Here, as for the original kernel $K(\omega,\tau)$, the limits 
$L\to \infty$ and $\omega \to 0$ of $\bar K$ do not commute, because
of the additional factor $L$ in $q$. Unfortunately, not only the small
but also the large $\omega$-dependence of the kernel is changed by the 
choice (\ref{aarts}). We arrive at a more adequate redefinition by 
first noting that for finite $L\,$ in the continuum
\be
\int\limits_0^{L} d\tau K(\omega,\tau) = \frac{2}{\omega}~, 
\label{knormc}
\ee
and on the lattice
\be
\sum\limits_{\tau=0}^{L-1} K(\omega,\tau) = \coth\frac{\omega}{2}~.
\label{knorml}
\ee
The choice 
\be
q(\omega)= \left\{ \begin{array}{c@{\qquad}l}
\omega/2 & {\rm in~the~continuum} \\ \tanh(\omega/2) & 
{\rm on~the~lattice}~, \end{array} \right.
\label{ourq}
\ee
normalizes then $\bar K(\omega,\tau)$ in $\tau$ to 1, independent of $L$, 
and leads to the sum rule
\be
\chi = \int\limits_0^{\infty} d\omega \bar A(\omega)~.
\label{sum}
\ee
This approach has several advantages: it cures the numerical problems
appearing in Bryan's algorithm, the limits of $\bar K$ for $L\to \infty$
and $\omega \to 0$ are finite and commute, and the large $\omega$-behaviour
remains unchanged on the lattice. From the measured $\chi$ the sum rule
gives us already some information on the size of $\bar A(\omega)$ 
before we start to calculate it. We note here, that the bosonic kernel
had been redefined earlier by Jarrel and Gubernatis \cite{Jarrell:1996} 
with a factor $\omega$ to ensure positive definite spectra.   

In applying the maximum entropy method a default model for the spectral
function is required which represents prior knowledge about $A(\omega)$
and accordingly $\bar A(\omega)\,$.
Quite obviously the transverse default model is always
\be
\bar A_T^D(\omega) = \chi_T \delta(\omega - m_T)~.
\label{deft} 
\ee 
The longitudinal case is more involved. In the high temperature 
region a single pole ansatz as for the transverse spectral function
seems to be adequate, since there for small $H$ all spin
components must show the same behaviour. For the low temperature 
phase we may derive a model spectral function directly from the 
considerations in the previous subsection. The corresponding lattice
version of the spectral function from Eqs.\ (\ref{dpp1}) and 
(\ref{dpp2}) is rather bulky. However, in the continuum limit
we can get a useful formula for $D_L(\tau)$ from the PP-relation
in the same way as we obtained Eq.\ (\ref{Dm}). The result is
\be
D_L(\tau) = \frac{N-1}{2M^2c_s^2}\;\frac{1}{L^2}\sum_{k_x,k_y}
\left[ {  e^{-m_k\tau}+e^{-m_k(L-\tau)} \over 2m_k (1 - e^{-m_k L}) }
\right]^2~,  \label{DLf1}
\ee
where $m_k^2= k_x^2+k_y^2+m_T^2\,$. This can be rewritten as
\be
D_L(\tau) = \frac{N-1}{2M^2c_s^2}\;\frac{1}{L^2}\sum_{k_x,k_y}
  \frac{1}{(2m_k)^2}
\left[\coth \left(\frac{m_kL}{2}\right)\frac{\bar K(2m_k,\tau)}{m_k}
 +\frac{1}{2\sinh^2 \left(\frac{m_kL}{2}\right)}
\right]~,
\label{DLf2}
\ee
which shows that $\bar A_L$ consists of $\delta$-function
contributions at $\omega=2m_k \ge 2m_T$ and at $\omega=0$. The 
latter disappears exponentially with increasing $L\,$.
The infinite volume expression is easily derived from Eq.\ (\ref{DLf1})
\ba
D_L(\tau)&\!\!\!\! = \!\!\!\!&\frac{N-1}{2M^2c_s^2}\;
\int\limits_{-\infty}^{\infty}\frac{d^2k}{(2\pi)^2}\frac{e^{-2m_k\tau}}
{4m_k^2} = \frac{N-1}{16\pi M^2c_s^2}\;
\int\limits_{m_T}^{\infty} dm_k \frac{e^{-2m_k\tau}}{m_k}~,\nonumber \\
&\!\!\!\! = \!\!\!\!&\frac{N-1}{16\pi M^2c_s^2}\;
\int\limits_{2m_T}^{\infty} d\omega \frac{e^{-\omega\tau}}{\omega}~.
\label{defl0}
\ea
Since the kernel reduces for $L\to\infty$ to
\be
\bar K(\omega,\tau) =\frac{\omega}{2} e^{-\omega \tau}~,
\label{kerninf}
\ee
we obtain a continuous spectrum starting at $\omega=2m_T$ and 
\be
\bar A_L^D(\omega) = \frac{N-1}{8\pi M^2c_s^2}\; \cdot
\frac{\theta(\omega-2m_T)}{\omega^2}~.
\label{defl}
\ee
The sumrule (\ref{sum}) leads again to the result (\ref{cl1}) for 
$\chi_L\,$.
%%%%%%%%%%%%%%%%%%%%%%%%%%%%%%%%%%%%%%%%%%%%%%%%%%%%%%%%%%%%%%%%%%%%%%%%%%%%%%%%

\section{Critical behaviour and scaling functions}
\label{section:Criti}

%%%%%%%%%%%%%%%%%%%%%%%%%%%%%%%%%%%%%%%%%%%%%%%%%%%%%%%%%%%%%%%%%%%%%%%%%%%%%%%%

In the thermodynamic limit ($V\rightarrow \infty$) the observables show
power law behaviour close to $T_c$. It is described by critical amplitudes
and exponents of the reduced temperature $t=(T-T_c)/T_c$ or magnetic field
$H\,$. The scaling laws at $H=0$ are 

\n for the magnetization 
\be
 M  \;=\; B (-t)^{\beta} \quad {\rm for~} t<0~,
\label{mcr}
\ee
the longitudinal susceptibility
\be
 \chi_L \;=\; C^{+} t^{-\gamma} \quad {\rm for~} t > 0~,
\label{chicr}
\ee 
and since for $H=0,t>0$ the correlation lengths coincide $\xi_T=\xi_L=\xi$ 
(like the susceptibilities)
\be
 \xi \;=\; \xi^{+} t^{-\nu} \quad {\rm for~} t > 0~.
\label{xicr}
\ee
On the critical line $T=T_c$ or $t=0$ we have for $H>0$ 
the scaling laws
\be
M \;=\; B^cH^{1/\delta} \quad {\rm or}\quad H \;=\;H_0 M^{\delta}~,
\quad H_0=(B^c)^{-\delta}~,
\label{mcrh}
\ee
\be
\chi_L \;=\; C^cH^{1/\delta-1} \quad {\rm with}\quad C^c \;=\;B^c/\delta~,
\label{chicrh}
\ee
and for the correlation lengths 
\be
\xi_{L,T} \;=\; \xi^c_{L,T} H^{-\nu_c}~,\quad \nu_c\;=\; \nu /\beta\delta~.
\label{xicrh}
\ee
We assume the following hyperscaling relations among the
critical exponents to be valid
\be
\gamma \;=\; \beta (\delta -1), \quad
d\nu \;=\; \beta (1 +\delta), \quad 2-\eta \;=\; \gamma/\nu~.
\label{hyps}
\ee
As a consequence only two critical exponents have to be known.
The dependence of the observables on the temperature and the external 
field in the critical region is described by scaling functions
\ba
M&\!\!\!\! = \!\!\!\!& h^{1/\delta}f_G(z)~,\quad \chi_L\; =\;
h^{1/\delta-1} f_{\chi}(z) /H_0~, \label{scf1}\\
\xi_T&\!\!\!\! = \!\!\!\!& h^{-\nu_c}g_{\xi}^T(z)~,\quad  
\xi_L\; =\;  h^{-\nu_c}g_{\xi}^L(z)~. \label{scf2} 
\ea
Here, the scaling variable $z$ is given by
\be
z=\bar t h^{-1/\Delta}~,\quad \Delta=\beta\delta~,
\label{zdef}
\ee
where $\bar t = t B^{1/\beta},~h=H/H_0$ are the normalized reduced
temperature and field. The normalization is chosen such that
\be
f_G(0)\; =\; 1~,\quad {\rm and}\quad f_G(z) {\raisebox{-1ex}{$
\stackrel{\displaystyle =}
{\scriptstyle z \rightarrow -\infty}$}}(-z)^{\beta}~.
\label{normfg}
\ee
The scaling functions $f_G(z)$ and $f_{\chi}(z)$ are universal and 
have been discussed in detail in Ref.\ \cite{Engels:2003nq}.
The functions  $g_{\xi}^{L,T}(z)$ are universal except for a normalization
factor. On the critical line $\bar t=0$ or $z=0$ we find from
(\ref{xicrh})
\be
g_{\xi}^{L,T} (0) \; =\;\xi^c_{L,T} (B^c)^{\nu/\beta}~,
\label{gxi0}
\ee
and from (\ref{xicr}) the asymptotic behaviour at $z \to \infty$,
that is for  $H\to 0,~t>0$
\be
g_{\xi}^{L,T} (z) \; {\raisebox{-1ex}{$\stackrel 
{\displaystyle =}{\scriptstyle z \rightarrow \infty}$}} \;
\xi^+ B^{\nu/\beta} z^{-\nu}~.
\label{gxiasy}
\ee
The ratios of the amplitude for $z\to \infty$ in (\ref{gxiasy}) 
and the $g_{\xi}^{L,T} (0)$ are universal and therefore also the
normalized scaling functions 
\be
{\hat g}_{\xi}^{L,T}(z) \; = \; g_{\xi}^{L,T}(z)/g_{\xi}^{L,T}(0)~.
\label{gnorm}
\ee
The stiffness $c_s$ in the critical region may be derived from Eq.\
(\ref{mt}), using $m_T=1/\xi_T$ and the scaling functions
\be
c_s = \frac{H}{M}h^{-2\nu_c}\left( g_{\xi}^T (z)\right)^2
= H_0\, h^{-\eta\nu_c}\;\frac{\left( g_{\xi}^T (z)\right)^2}{f_G(z)}~.
\label{cscrit}
\ee
The leading term of $c_s$ in the high temperature region $H\to 0,~t>0$
can be calculated in terms of critical amplitudes from Eq.\ (\ref{gxiasy})
and the corresponding equation for $f_G(z)$ \cite{Engels:2003nq}
\be
f_G (z)\; {\raisebox{-1ex}{$\stackrel 
{\displaystyle =}{\scriptstyle z \rightarrow \infty}$}}
\;  C^+ H_0 B^{\delta-1} z^{-\gamma}\;=\;R_{\chi} z^{-\gamma} ~.
\label{fcasy}
\ee
We find the simple result
\be
c_s = \big( \xi^+ \big)^2 \big( C^+ \big)^{-1} \;t^{-\eta\nu}\cdot 
[ 1+ {\cal O}\big( t^{-2\Delta} H^2\big) ]~.
\label{csasyp}
\ee
Here, the higher terms are proportional to $H^2$ because $\xi_T$ is an
even and $M$ an odd function of $H$ \cite{Engels:2003nq,Engels:2003xp}.
Therefore, $c_s$ is given for small
$H$ essentially by the leading term and its value is close to one.
In the low temperature phase both $\,\xi_T$ and $\chi_T$ diverge because
of the Goldstone effect. A corresponding analysis leads to the form
\be
c_s \sim (-t)^{-\eta\nu}\cdot[ 1 +{\cal O}\bigm( (-t)^{-\Delta/2}
H^{1/2}\bigm) ]~,
\label{csasym}
\ee
and here for small $H$ the higher terms cannot be neglected, they even
change the temperature dependence of $c_s$. We note, that the stiffness
$\rho_s$ defined by Privman et al.\ \cite{Privman:1991} is related to
our definition by $\rho_s=M^2 c_s$.   
 
%%%%%%%%%%%%%%%%%%%%%%%%%%%%%%%%%%%%%%%%%%%%%%%%%%%%%%%%%%%%%%%%%%%%%%%%%%%%%%%%

\section{Numerical details}
\label{section:Details}

%%%%%%%%%%%%%%%%%%%%%%%%%%%%%%%%%%%%%%%%%%%%%%%%%%%%%%%%%%%%%%%%%%%%%%%%%%%%%%%%
\n All our simulations were done on three-dimensional lattices with
periodic boundary conditions and linear extension $L=120$. As in Ref.\ 
\cite{Engels:2003nq} we have used the Wolff single cluster algorithm
\cite{Wolff:1988uh}, which was modified to include a non-zero magnetic 
field \cite{Dimitrovic:yd}. The coupling constant region which we have
explored was $0.90 \le J \le 1.2$, the magnetic field was varied from 
$H=0.0001$ to $H=0.007\,$.  We performed 150-400 cluster updates
between two measurements, such that the integrated autocorrelation time 
of the magnetization was typically of the order of 1. Only at the smallest 
$H-$values $\tau_{int}$ reached eventually 5. In general we 
made 100000 measurements at each fixed $H$ and $J$. Besides $M$ we 
measured the susceptibilities $\chi_{L,T}$ and the correlation functions 
$\langle S^{\parallel}(0) S^{\parallel}(\tau) \rangle$ and
$\langle {\v S}^{\perp}(0){\v S}^{\perp}(\tau) \rangle$ of the spin plane
averages for the plane distances $\tau=0,1,\dots,L/2$. Before the start
of the measurements 15000-20000 cluster updates were carried out to 
thermalize the system. In addition, up to 5000 of the first measurements
were discarded as long as the average cluster size was still rising.
In order to save storage we grouped the correlation function results in
blocks of 500 measurements. The block averages led thus to 190-200 sets 
of correlation data for the transverse and the longitudinal spins 
respectively. In the latter case we completed the data to connected 
correlation functions using the magnetization averages of the 
corresponding block of updates. These data were finally used as input
to our MEM-program.

\n In the following we are using $J_c=T_c^{-1}=0.93590$ from
Ref.\ \cite{Oevers:1996dt} as a value for the critical coupling.
%%%%%%%%%%%%%%%%%%%%%%%%%%%%%%%%%%%%%%%%%%%%%%%%%%%%%%%%%%%%%%%%%%%%%%%%%%%%%%%%

\subsection{The application of MEM}
\label{section:trymem}

%%%%%%%%%%%%%%%%%%%%%%%%%%%%%%%%%%%%%%%%%%%%%%%%%%%%%%%%%%%%%%%%%%%%%%%%%%%%%%%%
\n As already indicated in Section \ref{section:specfu} we determine
the spectral functions with Bryan's MEM-algorithm \cite{Bryan:1990} 
and the modified kernel $\bar K(\omega,\tau)$ where the lattice version of 
$q(\omega) = \tanh(\omega/2)$ is used. The Shannon entropy term which
contains the prior infor\-mation is then
\be
S= \int\limits_0^{\infty}d\omega \left[\bar A(\omega)-\bar A^D(\omega)
-\bar A(\omega) \ln( \bar A(\omega)/\bar A^D(\omega))\right]~,
\label{shanen}
\ee
where $\bar A^D(\omega)$ represents the default model. The MEM-algorithm
minimizes the 'free energy'
\be
F = \chi^2/2 -\alpha S~,
\label{freee}
\ee
or equivalently maximizes the conditional probability 
$P(\bar A|D \bar A^D) \sim \exp(-F)$. Here, $\chi^2/2$ is the logarithm
of the likelihood function for the data $D$ (assuming a Gaussian 
distribution of the data)
 and $\alpha$ is a parameter
balancing the relative importance of the data and prior knowledge. 
Bryan's algorithm calculates the spectral function at fixed $\alpha$ in 
a whole range of $\alpha$-values and estimates each time a conditional
probability $P(\alpha|D\bar A^D)$, which is typically sharply peaked
at some point $\hat\alpha$. The final result for $\bar A(\omega)$ is
then obtained by integrating over $\alpha$ with the conditional 
probability as weight. Further details of the MEM data analysis are
described in Refs.\ \cite{Asakawa:2000tr} and \cite{Bryan:1990}.
Our MEM program is based on an earlier version which was written
by Ines Wetzorke and was already used in Ref.\ \cite{Datta:2003ww}.
 
\begin{figure}[t]
\begin{center}
   \epsfig{bbllx=44,bblly=135,bburx=630,bbury=554,
       file=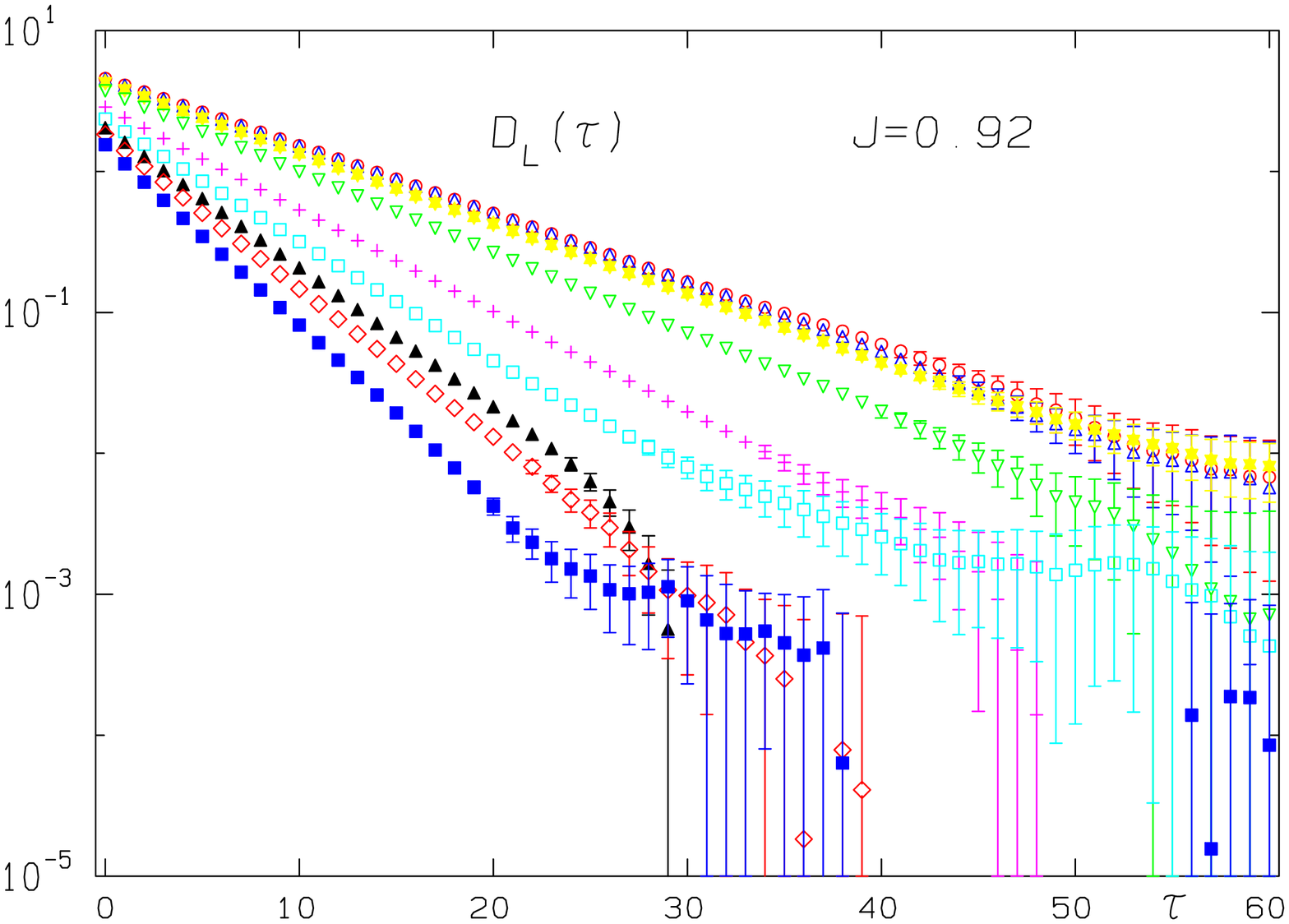, width=124mm}
\end{center}
\caption{The correlator $D_L(\tau)$ at $J=0.92$ for $H=0.0001$ (circles),
0.0002 (triangles), 0.0005 (davidstars), 0.001 (triangles down), 
0.002 (pluses), 0.003 (boxes), 0.004 (filled triangles),
0.005 (diamonds) and 0.007 (filled boxes).}
\label{fig:cl092}
%\end{figure}
%------------------------------------------------------------------------
%------------------------------------------------------------------------
%\begin{figure}[p!]
\begin{center}
   \epsfig{bbllx=44,bblly=135,bburx=630,bbury=554,
       file=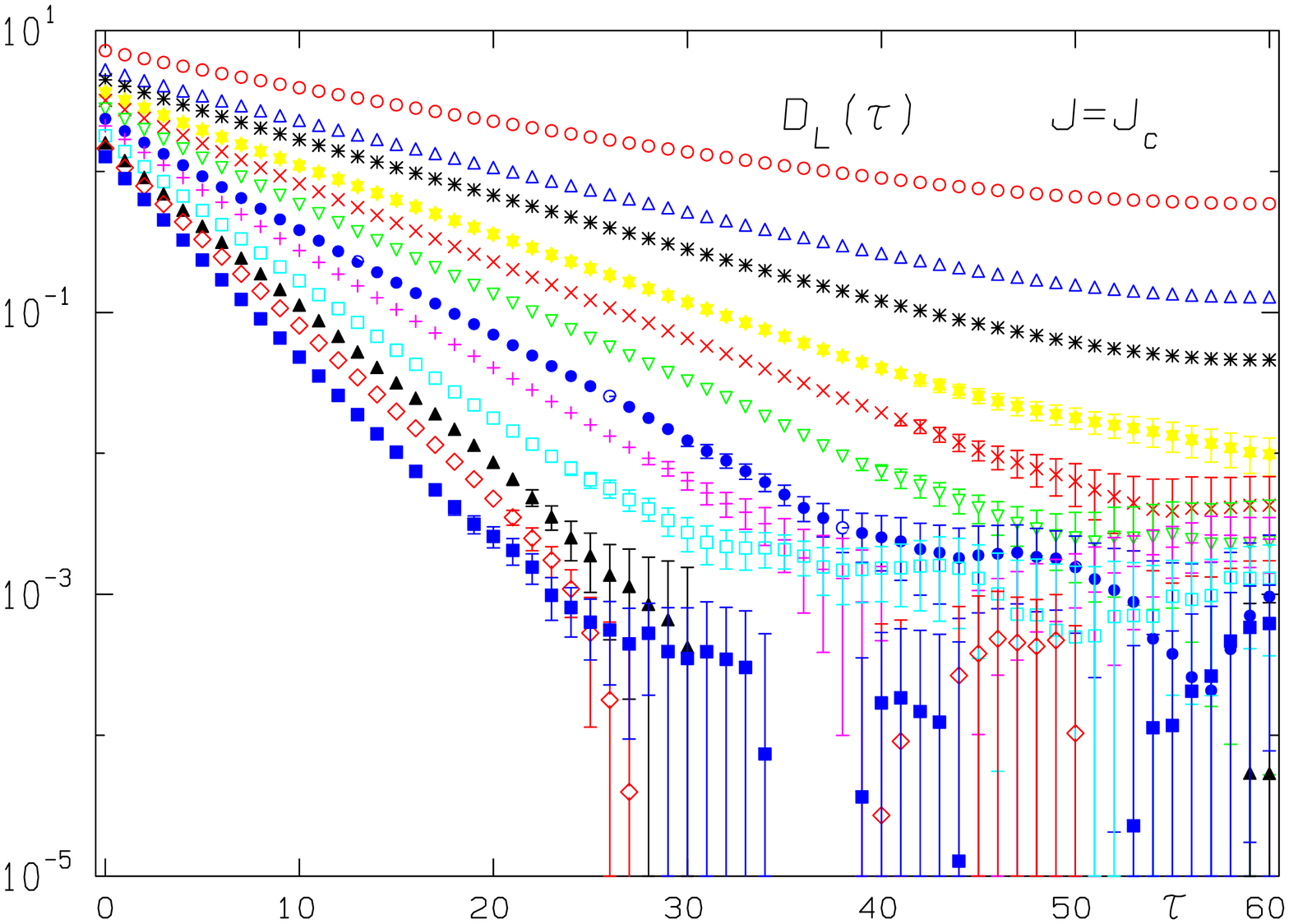, width=124mm}
\end{center}
\caption{The correlator $D_L(\tau)$ at $J=J_c$ for
the same $H$-values as in Fig.\ \ref{fig:cl092} and in addition for
$H=0.0003$ (stars), 0.0007 (crosses) and 0.0015 (filled circles).}
\label{fig:cljc}
\end{figure}

\begin{figure}[t]
\begin{center}
   \epsfig{bbllx=44,bblly=135,bburx=630,bbury=554,
       file=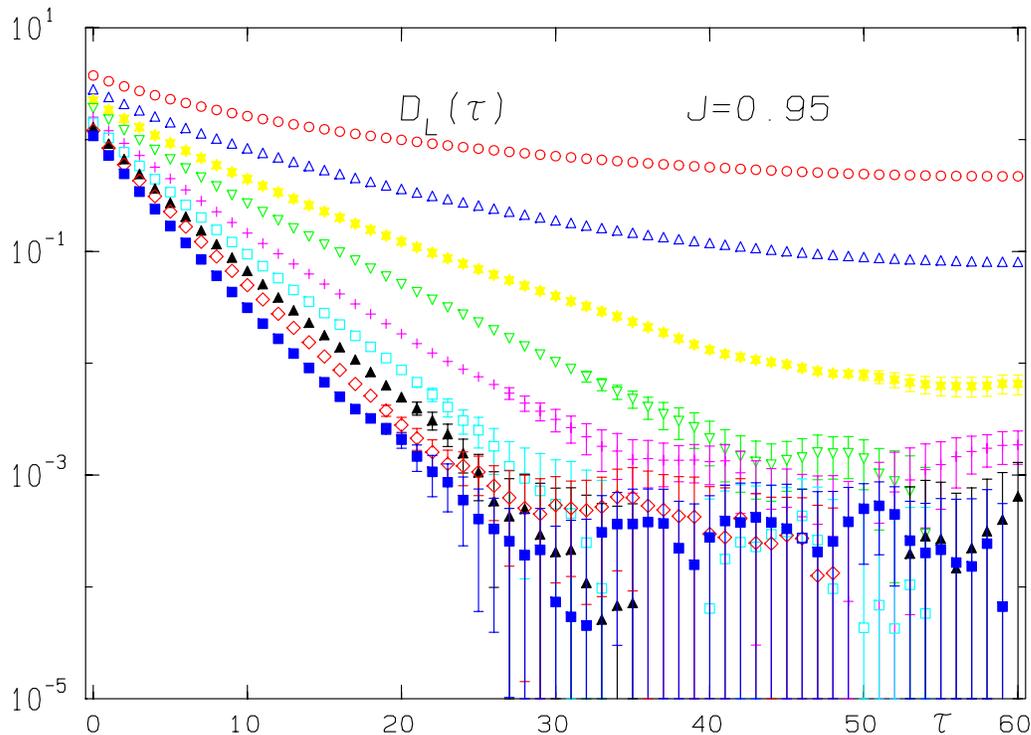, width=125mm}
\end{center}
\caption{The correlator $D_L(\tau)$ at $J=0.95$ for the same $H$-values 
and with the same notation as in  Fig.\ \ref{fig:cl092}.}
\label{fig:cl095}
\end{figure}

\n Due to the large number of $J$ and $H$ values, where we simulated the
$O(4)$-model, namely 95, we obtained rather different correlators.
In order to give an impression of our data we show the correlators 
$D_L(\tau)$ for the longitudinal sector at $J=0.92,~J_c$ and $J=0.95$
and all $H$-values in Figs.\ \ref{fig:cl092}, \ref{fig:cljc} and
\ref{fig:cl095}, that is one plot each for $T>T_c$, at $T_c$ and for 
$T<T_c\,$. The correlation functions were calculated from all data
belonging to the same $T$ and $H$ pair using the jackknife method. 
The results coincided within the errorbars with the averages from 
our MEM-input data sets.     
We observe at all couplings for increasing $H$ an increasing  
descent of the correlators and accordingly a shorter range of $\tau$
where the data are precise enough to be relevant for the MEM-analysis.
Whereas in the high temperature region the logarithm of the 
correlator is essentially a straight line, that is we have a sharp
single peak in the spectral function, the behaviour is gradually
changing with $T$ to a more complex spectral function in the low
temperature domain.

\n We proceed in the following way. Before we start with the MEM
analysis of a transverse or longitudinal correlator we check whether
it can be fitted with a one-pole ansatz, Eq.\ (\ref{dlat}) (in the 
longitudinal case evidently with $\chi_L$ and $m_L$). If the fit is
not completely satisfying we start MEM by choosing a default model,
eventually peaked at the previously found pole position. The suitable
form of the default model
\clearpage
%------------------------------------------------------------------------
%------------------------------------------------------------------------
\clearpage
\n  as such is not the only important 
requirement for a sensible result of the analysis.
The average size of the default model is also of significance. We
achieve this by normalizing the default model such that  
\be
\chi = \int\limits_0^{\omega_m} d\omega \bar A^D(\omega)~,
\label{Asize}
\ee
where $\omega_m$ is the maximal value of $\omega$ we use in MEM. 
If the integral in the last equation is definitely different from $\chi$
then the form of the default model is irrelevant for the MEM result, we
shall find only solutions for small $\alpha$-values with a peak at
$\hat \alpha$ in the range of 0.5 to 10 and a spectral function consisting
of isolated sharp peaks. Indeed, MEM performs then essentially only a 
$\chi^2$-fit, because the entropy term is suppressed by the small 
$\alpha$. The choice of the $\omega$-range is also of importance. 
%------------------------------------------------------------------------
%------------------------------------------------------------------------
Here we should keep in mind, that the spectral function at small $\omega$
determines the behaviour of the correlator at large $\tau$ and vice versa
the large $\omega$-region influences the correlator at small $\tau$. If
the selected maximal value $\omega_m$ is too small MEM delivers
nevertheless a spectral function which leads to a good correlator fit,
however at the expense of a spurious sharp peak in the spectral function
at $\omega_m$ and a distortion of the peaks at lower $\omega$. 
Alternatively, one may omit the lowest $\tau$-values from the analysis,
but then no information for large $\omega$ is obtained. Therefore we just
extend the $\omega$-range till the peak at the end point disappears. 
We choose the $\alpha$-range in such a way that the conditional probability
$P(\alpha|D\bar A^D)$ at the lower and upper interval limit is about
1\% of the value at the maximum position $\hat\alpha$. Finally we check 
the MEM result for the spectral function by reconstructing the correlator
and comparing the result with the correlator data. 
%%%%%%%%%%%%%%%%%%%%%%%%%%%%%%%%%%%%%%%%%%%%%%%%%%%%%%%%%%%%%%%%%%%%%%%%%%%%%%%%

\section{Results for the spectral functions}
\label{section:Results}

%%%%%%%%%%%%%%%%%%%%%%%%%%%%%%%%%%%%%%%%%%%%%%%%%%%%%%%%%%%%%%%%%%%%%%%%%%%%%%%%

\subsection{The transverse spectrum}
\label{section:transspec}

\begin{figure}[b]
\begin{center}
   \epsfig{bbllx=25,bblly=135,bburx=611,bbury=552,
       file=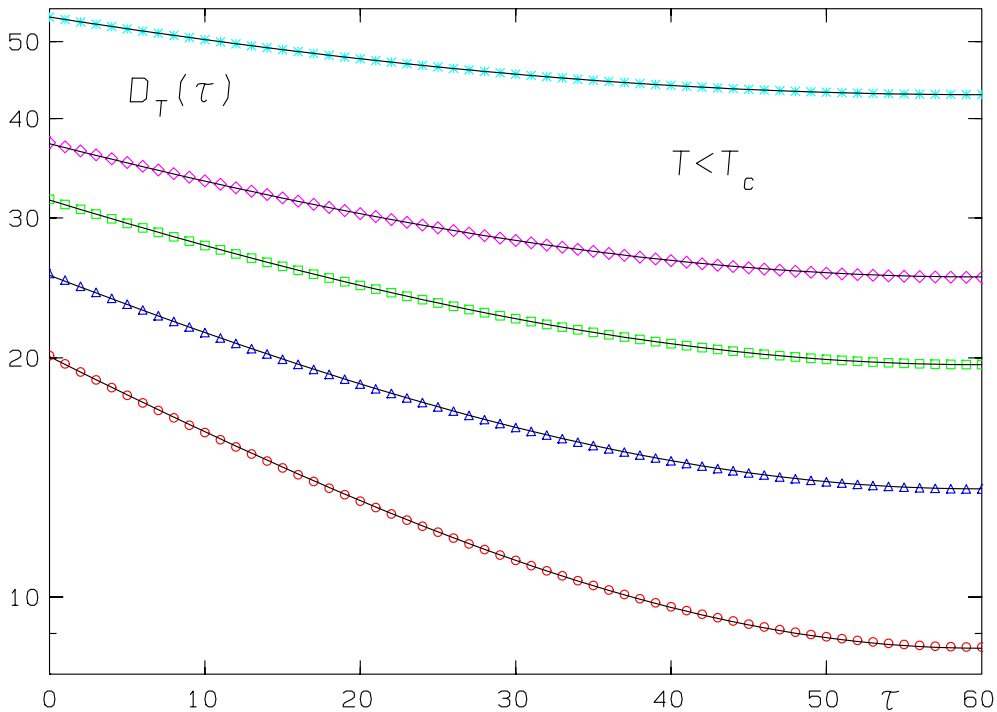, width=124mm}
\end{center}
\caption{The correlators $D_T(\tau)$ for $H=0.0001$ below $T_c$
at $J=0.94$ (circles), 0.95 (triangles), 0.97 (boxes), 1.0 (diamonds) 
and 1.2 (stars). The lines are the respective fits with the Gaussian 
form, Eq.\ (\ref{dlat}).}
\label{fig:tlttc}
\end{figure}

\begin{figure}[t]
\begin{center}
   \epsfig{bbllx=32,bblly=135,bburx=618,bbury=552,
       file=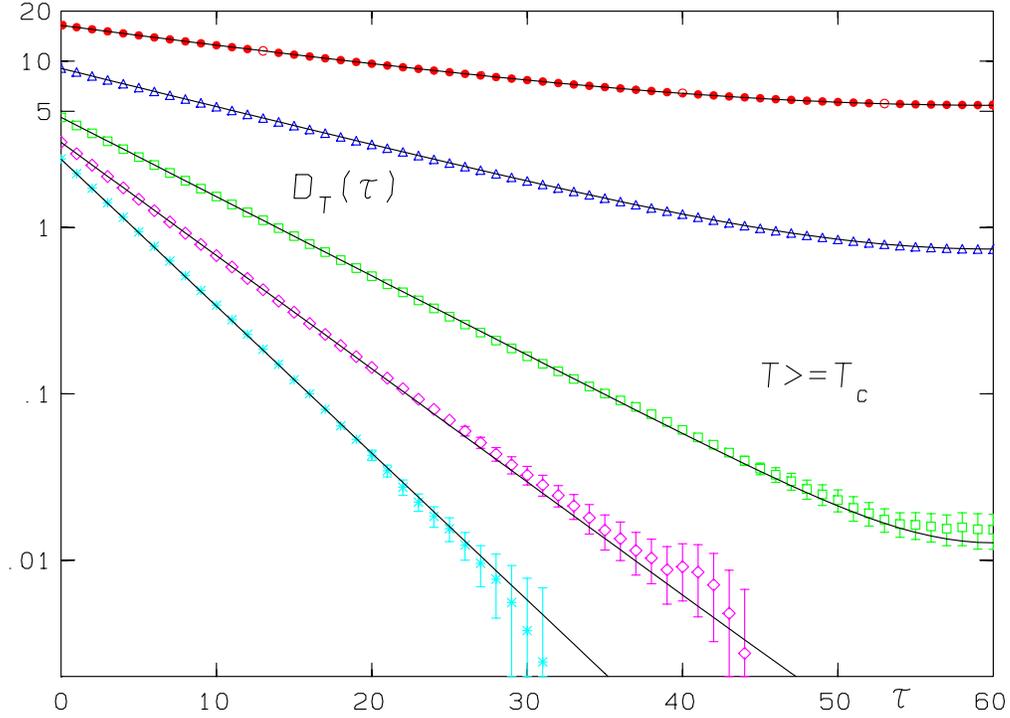, width=124mm}
\end{center}
\caption{The correlators $D_T(\tau)$ for $H=0.0001$ at $T_c$
(filled circles) and above $T_c$ at $J=0.93$ (triangles), 0.92 (boxes),
0.91 (diamonds) and 0.90 (stars). The lines are the respective fits with
the Gaussian form, Eq.\ (\ref{dlat}).}
\label{fig:tgttc}
\end{figure}

%%%%%%%%%%%%%%%%%%%%%%%%%%%%%%%%%%%%%%%%%%%%%%%%%%%%%%%%%%%%%%%%%%%%%%%%%%%%%%%%
As a first check we have tried a fit of our transverse correlator data 
with the Gaussian form, Eq.\ (\ref{dlat}). The value for $\chi_T$
required in the equation, was either calculated from Eq.\ (\ref{cmoh}) 
or Eq.\ (\ref{fluc}). Also, the $\tau$-range for the
fit was varied to obtain sensible results. In {\em all} cases, that is 
for all our $J$ and $H$-values, the transverse correlator data could 
be represented by the Gaussian one-pole form, Eq.\ (\ref{dlat}).
As examples we show in Figs.\ \ref{fig:tlttc} and \ref{fig:tgttc} our
transverse correlator data for $H=0.0001$. In Fig.\ \ref{fig:tlttc}
all data for $T<T_c$ are shown, in Fig.\ \ref{fig:tgttc} the data 
at $T_c$ and for $T>T_c$. Below $T_c$ the $m_T$-values for $H=0.0001$
are rather small so that the correlators are comparatively flat, the
overall size, being proportional to $\chi_T=M/H$, diminishes when $T_c$
is approached
%------------------------------------------------------------------------
%------------------------------------------------------------------------
\clearpage
%------------------------------------------------------------------------
%------------------------------------------------------------------------
\n from below. Above $T_c$ the average size diminishes further with 
increasing $T$, however the $m_T$-value increases (the correlation length
$\xi_T=1/m_T$ decreases) such that the correlators become steeper.

\begin{figure}[p!]
\begin{center}
   \epsfig{bbllx=41,bblly=151,bburx=632,bbury=569,
       file=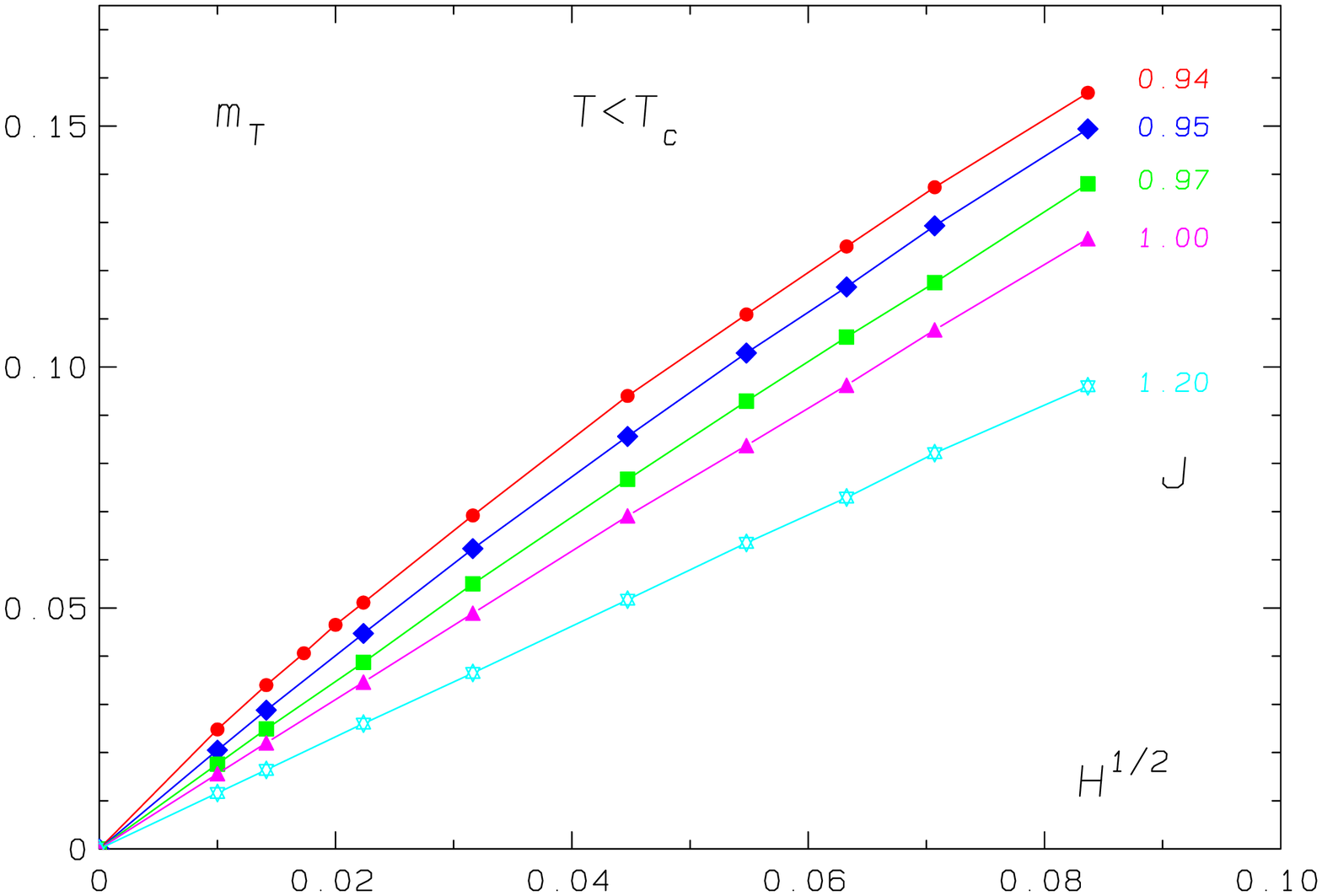, width=122mm}
\end{center}
\caption{The transverse mass $m_T$ as a function of $H^{1/2}$ in the 
low temperature phase $T<T_c$. The $J$-values are 0.94 (circles),
0.95 (diamonds), 0.97 (squares), 1.0 (triangles) and 1.2 (davidstars).}
\label{fig:mtcold}
%\end{figure}
%------------------------------------------------------------------------
%------------------------------------------------------------------------
%\begin{figure}[b]
\begin{center}
   \epsfig{bbllx=36,bblly=135,bburx=620,bbury=553,
       file=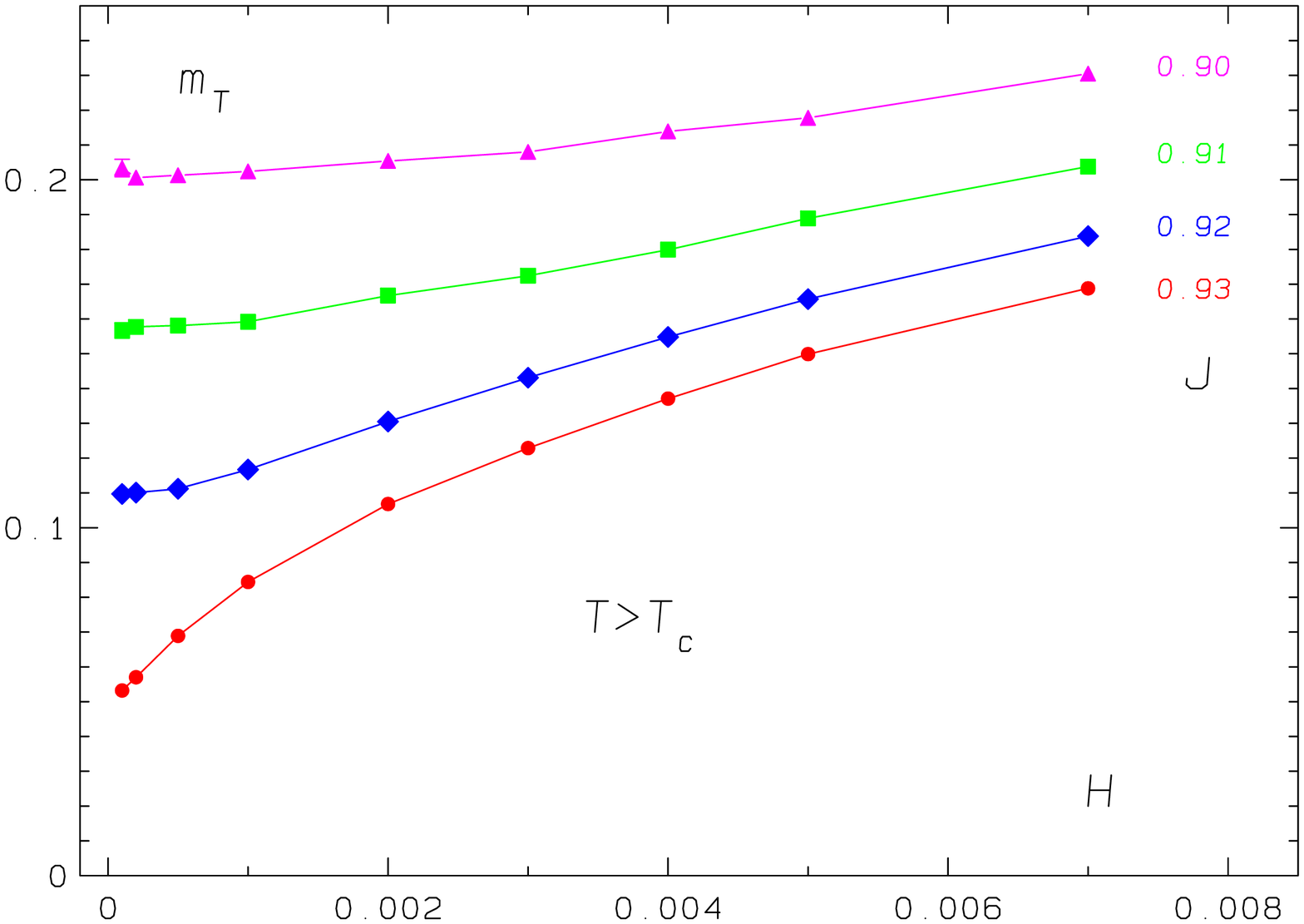, width=122mm}
\end{center}
\caption{The transverse mass $m_T(H)$ for $T>T_c$. The $J$-values are
 0.93 (circles), 0.92 (diamonds), 0.91 (squares) and 0.90 (triangles).}
\label{fig:mthot}
\end{figure}

\begin{figure}[t]
\begin{center}
   \epsfig{bbllx=46,bblly=192,bburx=632,bbury=527,
       file=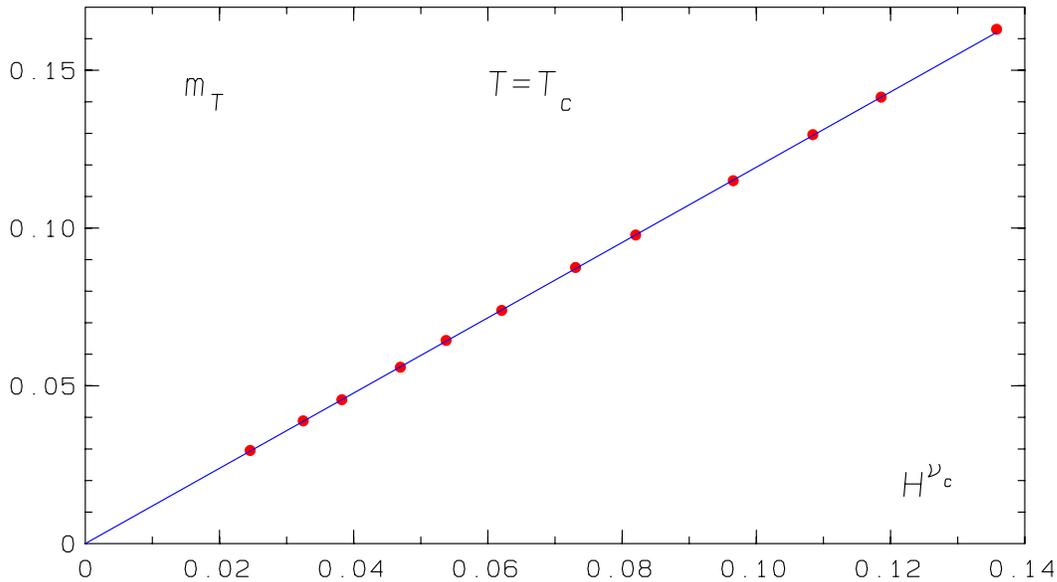, width=125mm}
\end{center}
\caption{The transverse mass $m_T$ (circles) versus $H^{\nu_c}$ on the 
critical line. The straight line is given by Eq.\ (\ref{mtcrit}) with 
the parameters from Ref.\ \cite{Engels:2003nq}.}
\label{fig:mtjc}
\end{figure}

\n In Figs.\ \ref{fig:mtcold} and \ref{fig:mthot} we show the results 
from the Gaussian fits for the transverse mass $m_T$ below and above $T_c$.
The data points for the same $J$-values have been connected by straight
lines to guide the eye.
Since we expect in the low temperature region near the coexistence line 
that $m_T\sim H^{1/2}$ (see Eq.\ (\ref{xigold})), we have plotted $m_T$
versus $H^{1/2}$ in Fig.\ \ref{fig:mtcold}. Indeed, we observe with
decreasing temperature an increasing range in $H^{1/2}$ with a straight
line behaviour. On the other hand, close to $T_c$, here for $J=0.94$, the
curvature of $m_T(H^{1/2})$ grows slightly. Above $T_c$ we expect for
$H\rightarrow 0$ a finite transverse correlation length and therefore 
that $m_T$ reaches a finite value. This expectation is borne out in 
Fig.\ \ref{fig:mthot}, where we have plotted $m_T$ as a function of $H$.  
As in the low temperature range the curvature of $m_T(H)$ is growing
when $T_c$ is approached. The $H$-dependence of the transverse mass
on the critical line $T=T_c$ results from Eq.\ (\ref{xicrh}) in
\be
m_T = {\xi_T^c}^{-1}\cdot H^{\nu_c}~.
\label{mtcrit}
\ee
In Fig.\ \ref{fig:mtjc} we show the $m_T$-data at $T_c$ as a function
of $H^{\nu_c}$, where we used the value $\nu_c=0.4024$ obtained in 
Ref.\ \cite{Engels:2003nq}. The line through the data points is given
by Eq.\ (\ref{mtcrit}) with $\xi_T^c=0.838$ again from 
\cite{Engels:2003nq}. The data fully agree with the expectations.
%%%%%%%%%%%%%%%%%%%%%%%%%%%%%%%%%%%%%%%%%%%%%%%%%%%%%%%%%%%%%%%%%%%%%%%%%%%
%\clearpage
%------------------------------------------------------------------------
%------------------------------------------------------------------------
\newpage
%------------------------------------------------------------------------
%------------------------------------------------------------------------
\n A straight line fit to the data for $\ln m_T$ as a function of $\ln H$
leads to the following results for the parameters
\be
\xi_T^c = 0.839\pm 0.004~,\quad {\rm and}\quad \nu_c = 0.4020\pm 0.0007~,
\label{resmt}
\ee
confirming the previous numbers.
%%%%%%%%%%%%%%%%%%%%%%%%%%%%%%%%%%%%%%%%%%%%%%%%%%%%%%%%%%%%%%%%%%%%%%%%%%

\subsection{The longitudinal spectrum}
\label{section:longspec}

%%%%%%%%%%%%%%%%%%%%%%%%%%%%%%%%%%%%%%%%%%%%%%%%%%%%%%%%%%%%%%%%%%%%%%%%%%
\n In the high temperature phase the longitudinal and transverse 
correlators must coincide for $H\rightarrow 0$ and therefore as well
their spectral functions, that is the correlators will both have the
same Gaussian form. With increasing $H$ and/or decreasing temperature
the two correlators will increasingly differ until the longitudinal
mass $m_L$ reaches $2m_T$, the threshold value of the longitudinal
continuum in the low temperature phase. Indeed, that is what we 
observe in our analysis.

\begin{figure}[p!]
\begin{center}
   \epsfig{bbllx=36,bblly=147,bburx=621,bbury=566,
       file=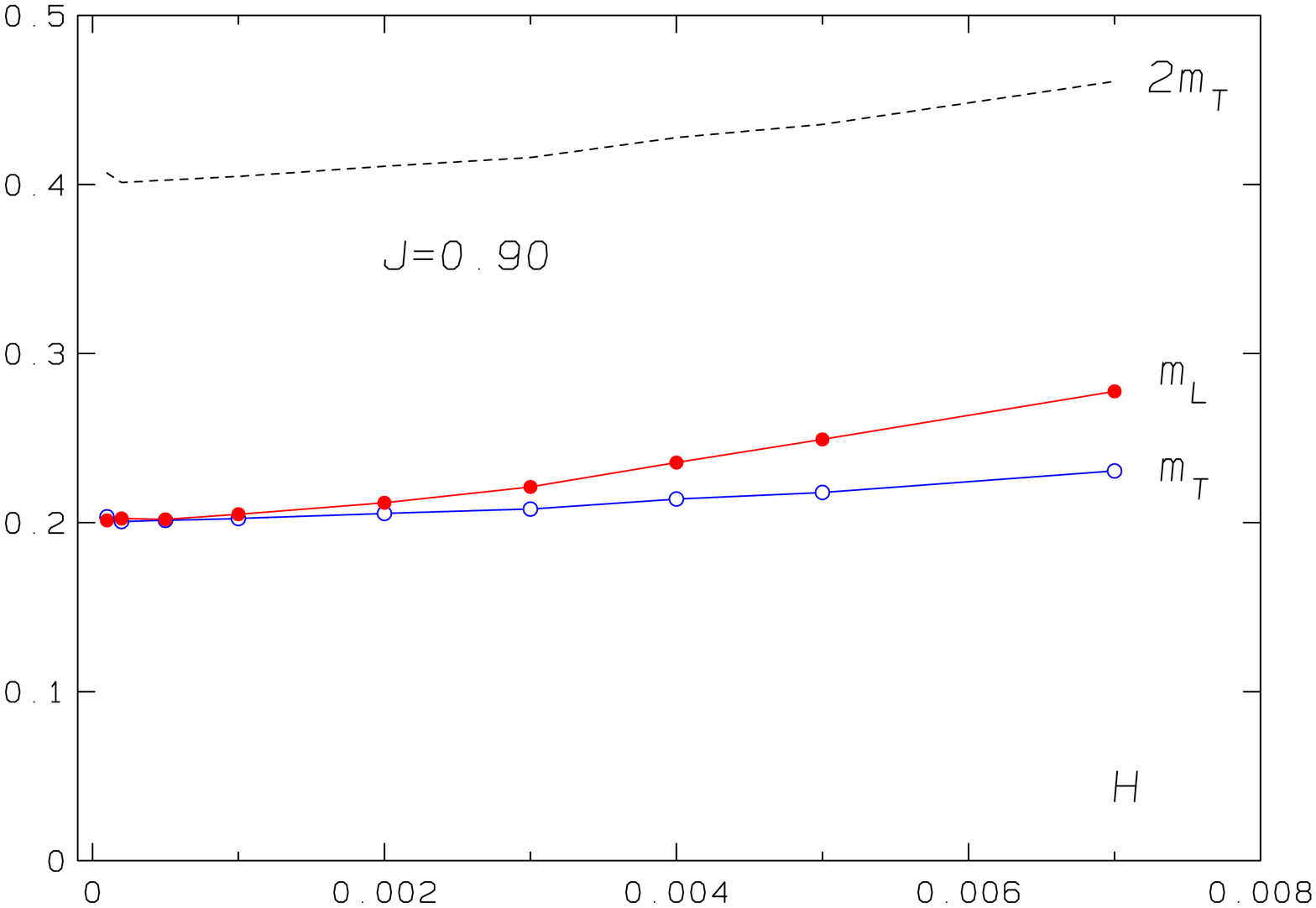, width=120mm}
\end{center}
\caption{The longitudinal mass $m_L(H)$ for $J=0.90$ (filled circles).
For comparison $m_T$ and $2m_T$ are also shown.}
\label{fig:ml090}
%\end{figure}
%------------------------------------------------------------------------
%------------------------------------------------------------------------
%\begin{figure}[b]
\begin{center}
   \epsfig{bbllx=36,bblly=150,bburx=621,bbury=568,
       file=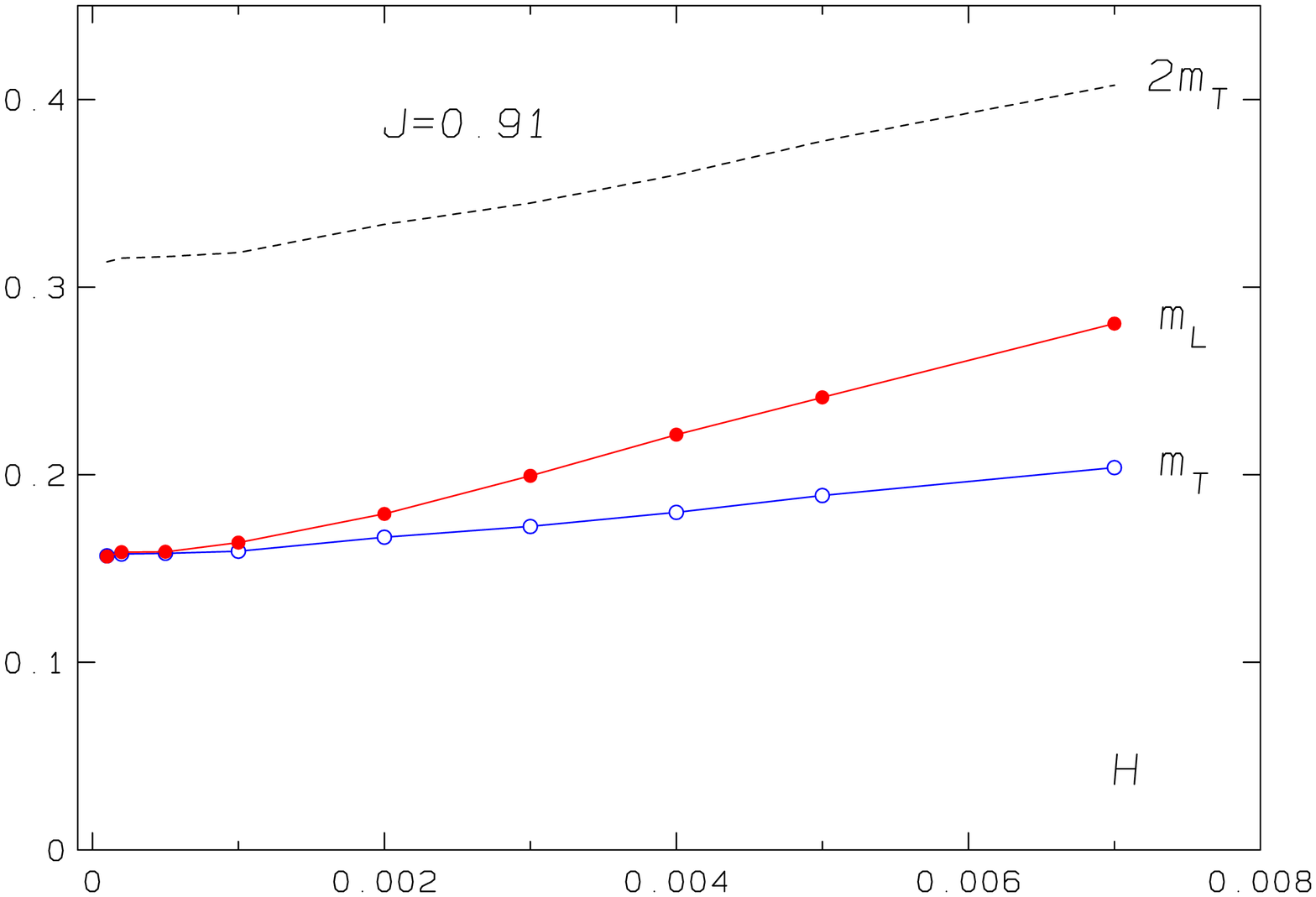, width=120mm}
\end{center}
\caption{The longitudinal mass $m_L(H)$ for $J=0.91$ (filled circles).
For comparison $m_T$ and $2m_T$ are also shown.}
\label{fig:ml091}
\end{figure}

\begin{figure}[b!]
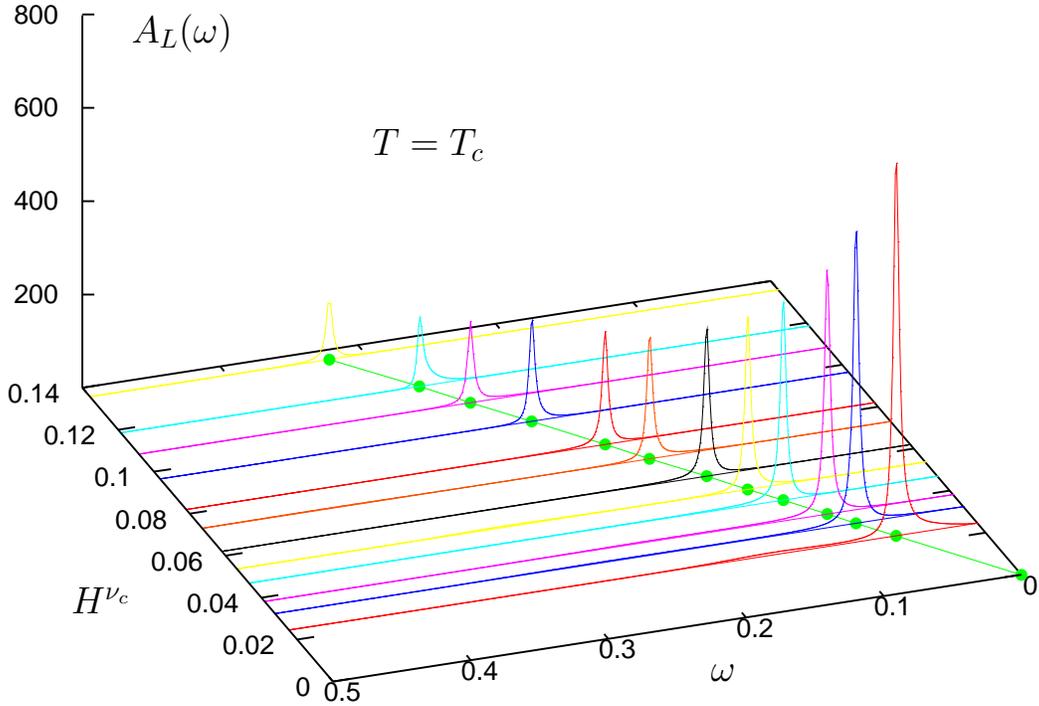

\caption{The spectral function $A_L(\omega)$ at $T_c$ and fixed values 
of $H$ plotted versus $H^{\nu_c}$. For comparison the values of $2m_T$
(filled circles) are also shown.}
\label{fig:jcal}
\end{figure}

\n At the two highest temperatures, corresponding to  $J=0.90$ and
0.91, we find that the longitudinal correlators for all our $H$-values
are of the Gaussian form, Eq. (\ref{dlat}). In Figs.\ \ref{fig:ml090}
and \ref{fig:ml091} we show the results for the longitudinal mass 
$m_L$ in comparison to the respective transverse mass $m_T$ and $2m_T$
as a function of $H$. We see that at the lowest $H$-value the two 
masses coincide and that $m_L$ becomes larger than
%------------------------------------------------------------------------
%------------------------------------------------------------------------
%------------------------------------------------------------------------
\begin{figure}[p!]
\begin{center}
   \epsfig{bbllx=36,bblly=147,bburx=621,bbury=566,
       file=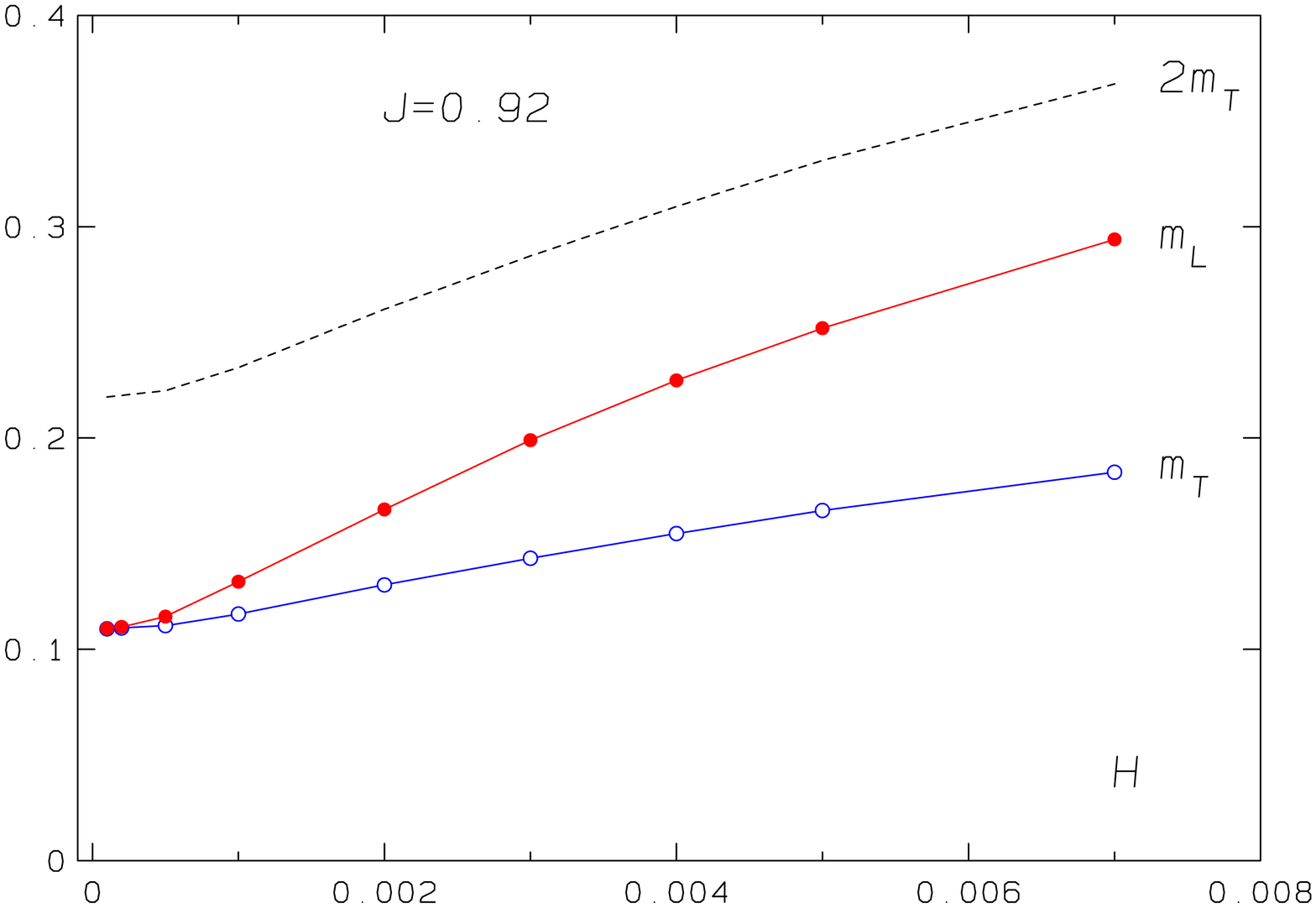, width=120mm}
\end{center}
\caption{The longitudinal mass $m_L(H)$ for $J=0.92$ (filled circles).
For comparison $m_T$ and $2m_T$ are also shown.}
\label{fig:ml092}
%\end{figure}
%------------------------------------------------------------------------
%------------------------------------------------------------------------
%\begin{figure}[b]
\begin{center}
   \epsfig{bbllx=36,bblly=147,bburx=621,bbury=566,
       file=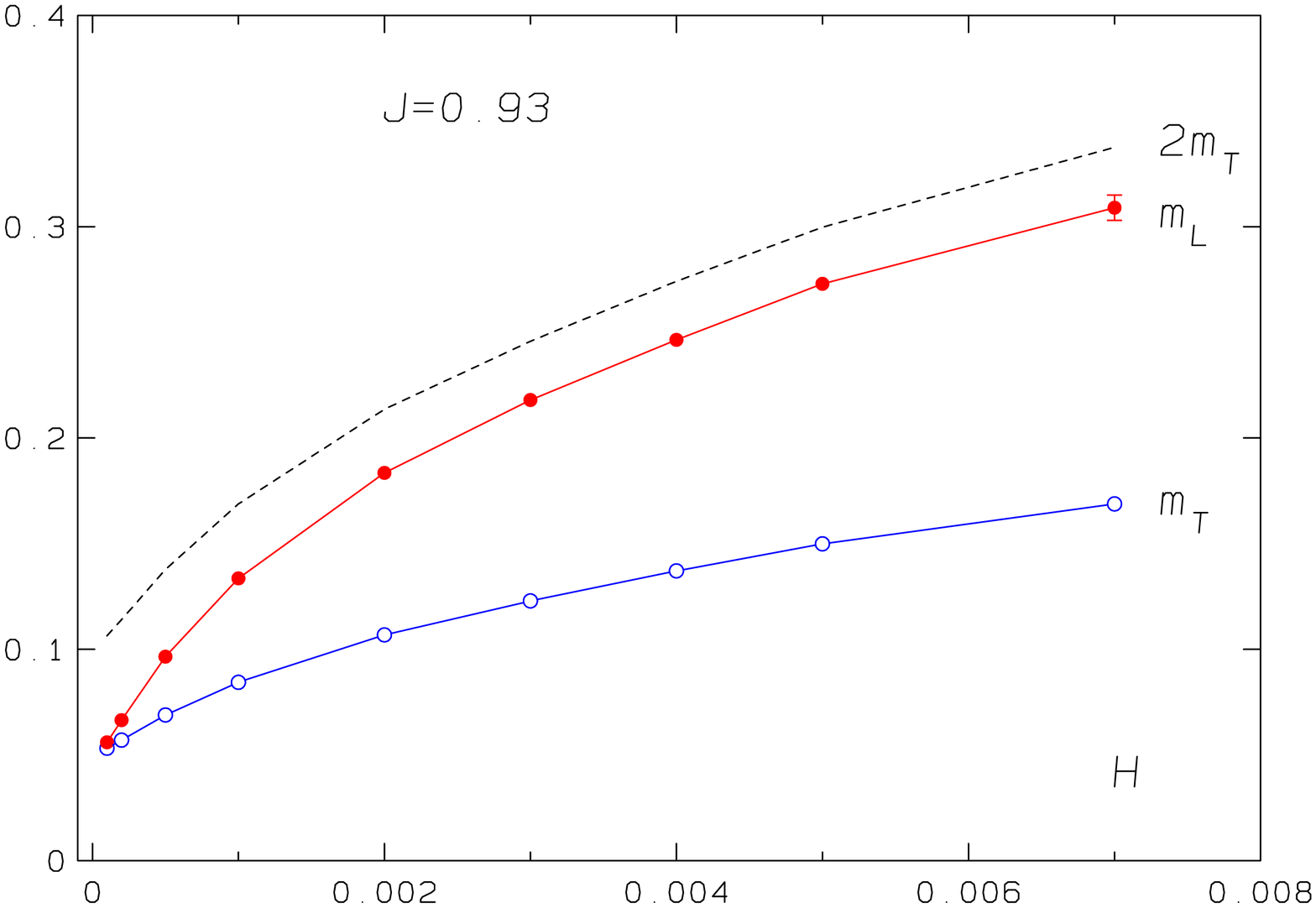, width=120mm}
\end{center}
\caption{The longitudinal mass $m_L(H)$ for $J=0.93$ (filled circles).
For comparison $m_T$ and $2m_T$ are also shown.}
\label{fig:ml093}
\end{figure}
%------------------------------------------------------------------------
$m_T$ with increasing external field. The relative difference increases 
with lowering temperature. In approaching $T_c$ further, at $J=0.92$ 
and 0.93, we still find a single peak in the spectral function, however
instead of a $\delta$-function contribution, the Gauss form, a slightly
better discription is given by a peak with a finite width. We have
therefore approximated the $\delta$-function by a Breit-Wigner form
\be
\delta_{BW} (\omega-m_L) = \frac{1}{2\pi}\cdot \frac{\Gamma}
{(\omega-m_L)^2 + (\Gamma/2)^2}~.
\label{dBW}
\ee
ls
r default model function for MEM is then given by
\be
\bar A^D_L(\omega)= \chi_L \delta_{BW} (\omega-m_L)~. 
\label{ADBW}
\ee
For $J=0.92$ and 0.93 we got the best results for a width
$\Gamma=0.005$, a maximal value $\omega_m=3$, and a resolution 
$\Delta\omega=0.001$ of the frequency range. Since the Breit-Wigner
form (\ref{dBW}) is normalized to 1 in the range $[-\infty,\infty]$
the cut-off at the limits
of the $\omega$-range $[0,\omega_m]$ was corrected by a corresponding
factor in $\bar A^D_L$. In Figs.\ \ref{fig:ml092} and \ref{fig:ml093} we
show the results for $m_L$. The further approach of $m_L$ to the value
of $2m_T$ is manifest, in particular in Fig.\ \ref{fig:ml093}. As 
expected from the result of Ref.\ \cite{Engels:2003nq}, where on the
critical line a ratio $\xi^c_T/\xi^c_L=1.99(1)$ was obtained, we find at
$T_c$ that $m_L$ coincides with $2m_T$. This is seen in Fig.\  
\ref{fig:jcal}, where we show the spectral functions $A_L(\omega)$ 
(not $\bar A_L(\omega)$) for each of our $H$-values. Again, we have used 
the default model of Eq.\ (\ref{ADBW}) with the width $\Gamma=0.005$ and 
the resolution $\Delta\omega=0.001$. The maximal value $\omega_m$ was 
in the range $2-4$. 

%----------------------------------------------------------------------------
\setlength{\unitlength}{1cm}
\begin{picture}(10,9.5)
\put(1,0){ 
   \epsfig{bbllx=498,bblly=163,bburx=146,bbury=657,
       file=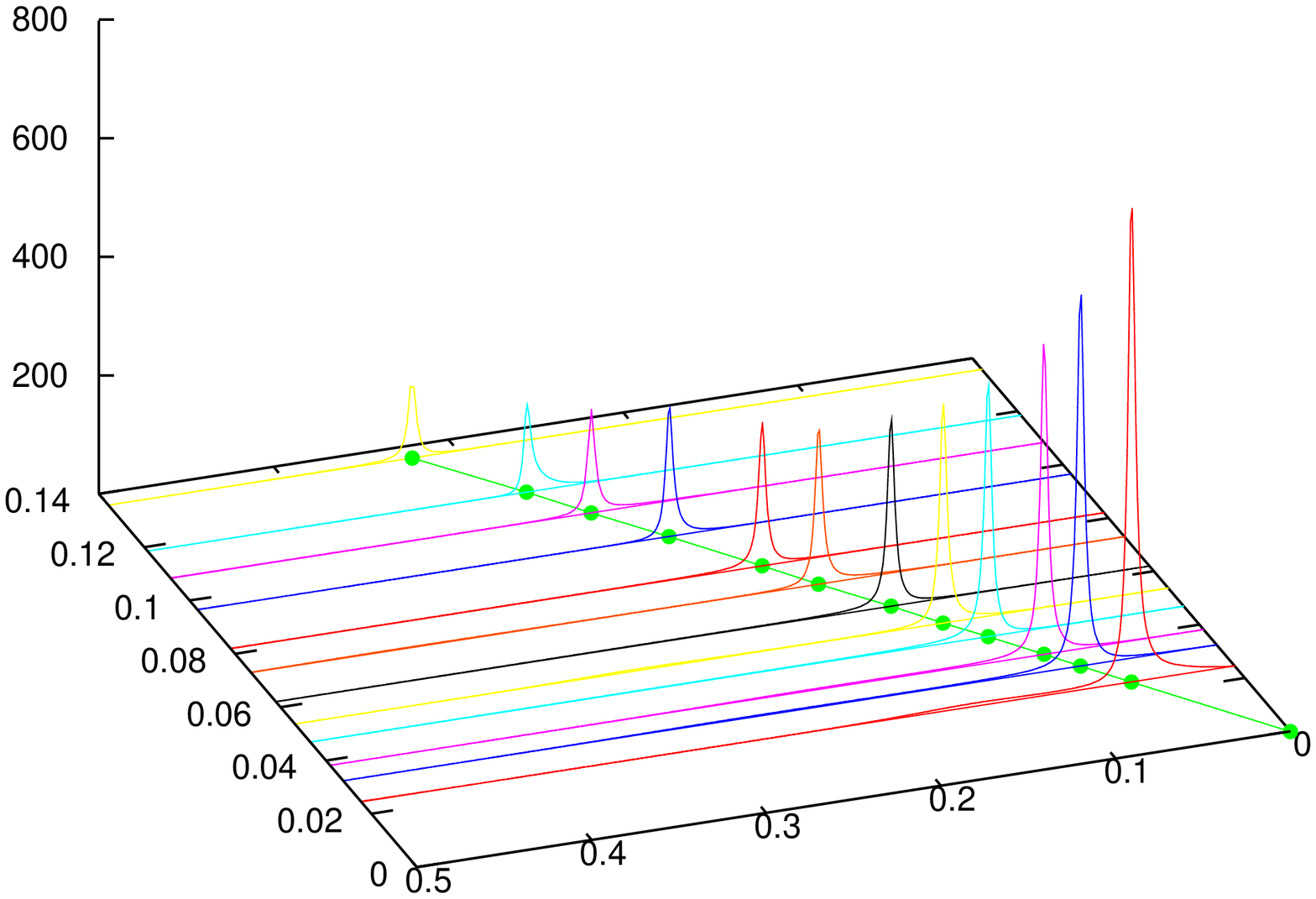,height=125mm,angle=-90}
          }
\put(1.8,8.5){\large{$A_L(\omega)$}}
\put(5.0,7.0){\large{$T=T_c$}}
\put(9.5,0){\large{$\omega$}}
\put(1,0.9){\large{$H^{\nu_c}$}}
\end{picture}
%------------------------------------------------------------------------
%------------------------------------------------------------------------
%------------------------------------------------------------------------
\newpage  
\n One can use as well a larger width of for example 0.01 to obtain
similarly good correlation functions. The peak heights shrink then
correspondingly. It is remarkable that no further peak or continuum
contributions appear in the spectral functions at the critical temperature.

\begin{figure}[b!]
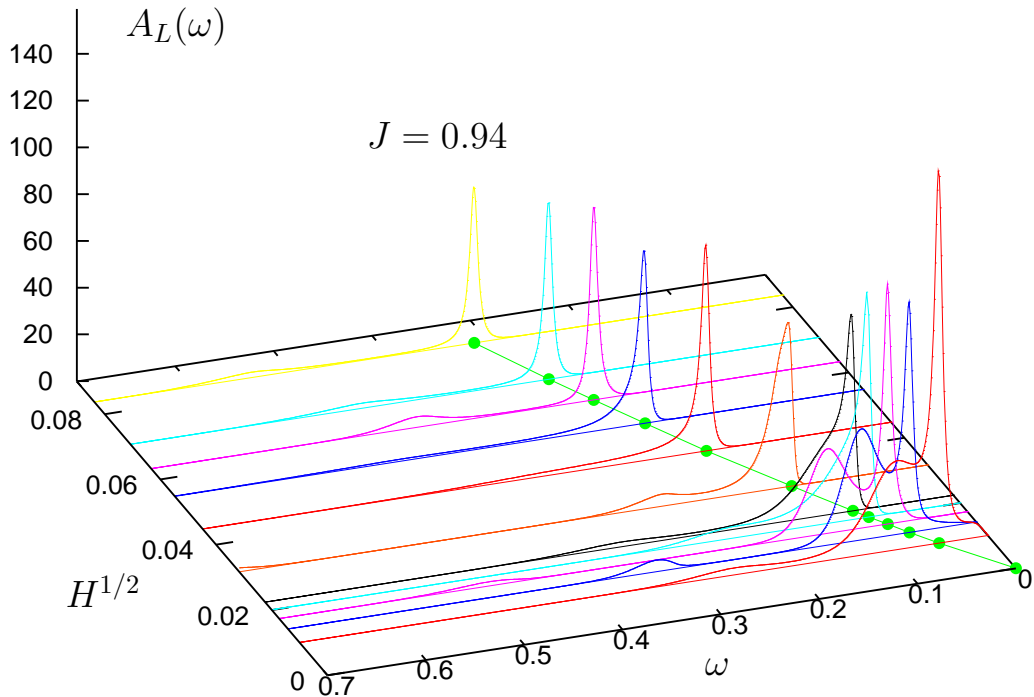

\caption{The spectral function $A_L(\omega)$ at $J=0.94$ and fixed values 
of $H$ plotted versus $H^{1/2}$. For comparison the values of $2m_T$
(filled circles) are also shown.}
\label{fig:ao094l}
\end{figure}

\n In the low temperature phase where we expect a continuum contribution 
in the longitudinal spectral function starting at the threshold value
$2m_T$ we have tried several forms for the default model. First
we have used the continuum form of Eq. (\ref{defl}). Here, the default 
spectral function is strictly zero below $\omega=2m_T$ and the MEM
result remains always zero there. We have therefore moderated the model
by combining the left half of a sharp Breit-Wigner form at the threshold
with the continuum form above. That allows for non-zero spectral 
function values also below the threshold. Yet, this default model is
predicting a too low threshold peak and too large contributions at high
$\omega$-values. It turned out that indeed a Breit-Wigner default model
centered at the threshold nevertheless allows for continuum contributions
well above its peak position and delivers the best results. In contrast,
a corresponding Gaussian distribution leads to poor results. Obviously,
the tail of the Gaussian distribution is dying out too fast, whereas the
tail of the Breit-Wigner form falls off with $\omega^{-2}$ just as the
continuum default model. In Fig.\ \ref{fig:ao094l} we show the
MEM results for the spectral functions $A_L(\omega)$ at $J=0.94$, the
temperature value
\vspace{0.7cm}
%----------------------------------------------------------------------------
\setlength{\unitlength}{1cm}
\begin{picture}(10,9.5)
\put(1,0){ 
   \epsfig{bbllx=498,bblly=163,bburx=146,bbury=657,
       file=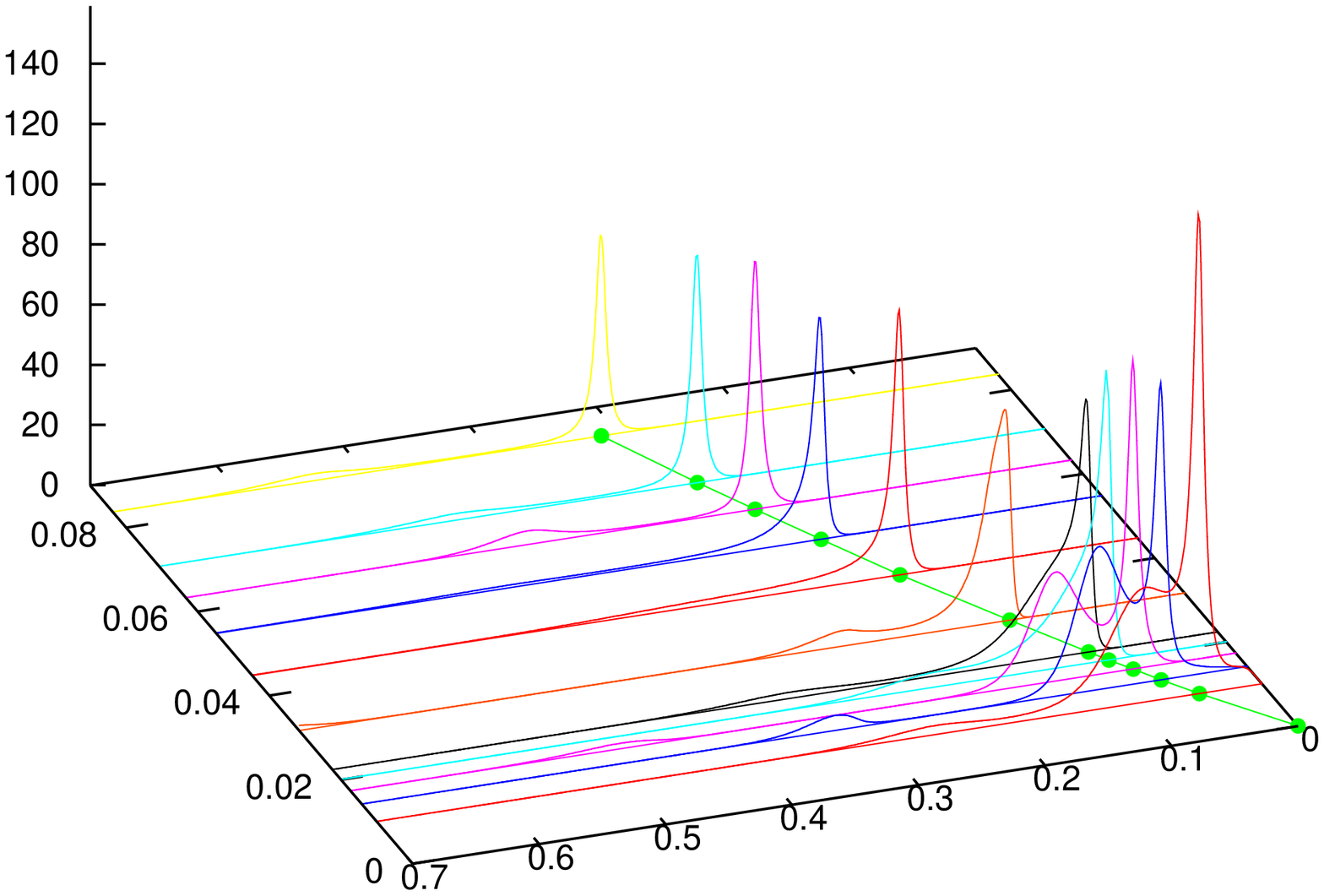,height=125mm,angle=-90}
          }
\put(1.8,8.5){\large{$A_L(\omega)$}}
\put(5.0,7.0){\large{$J=0.94$}}
\put(9.5,0){\large{$\omega$}}
\put(1,0.9){\large{$H^{1/2}$}}
\end{picture}
%------------------------------------------------------------------------
%------------------------------------------------------------------------
%------------------------------------------------------------------------
\newpage  
\n  in the cold phase closest to $T_c\,$. As in Fig.\
\ref{fig:mtcold} we plot the data versus $H^{1/2}$. The parameters
we used for MEM and the default model were $\Gamma=0.01$,  
$\Delta\omega=0.001$ and $\omega_m$ was again in the range $2-4$. We
observe that apart from the main peak at the threshold value $\omega=2m_T$
additional peaks appear at higher $\omega$-values, in particular for small
external fields $H$. In other respects the picture at $J=0.94$ resembles
still the one at $T_c$, Fig.\ \ref{fig:jcal}: the peaks at the threshold
are the major contributions to the spectral functions. 

\begin{figure}[h]
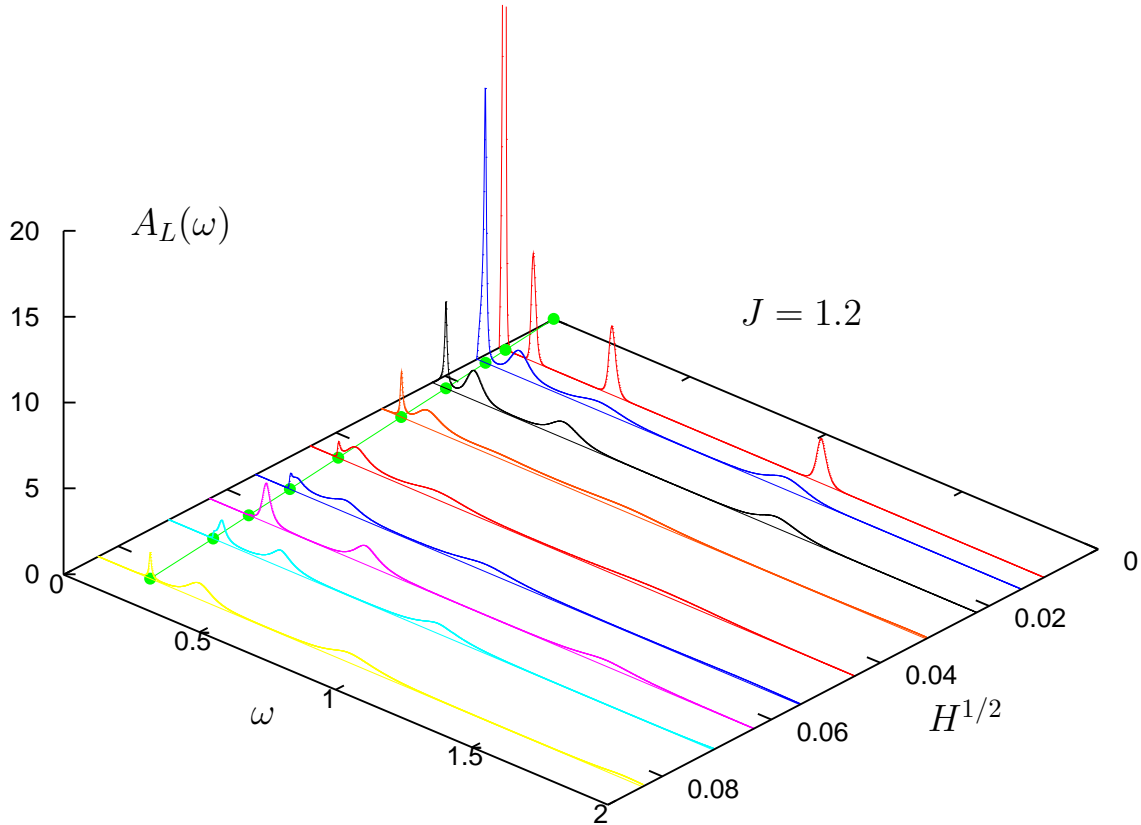

\caption{The spectral function $A_L(\omega)$ at $J=1.2$ and fixed values 
of $H$ plotted versus $H^{1/2}$. For comparison the values of $2m_T$
(filled circles) are also shown.}
\label{fig:ao12l}
\end{figure}

\n The behaviour of the spectral function changes gradually when the 
temperature is shifted deeper into the cold phase. Apart from the smallest
$H$-values the continuum contributions become more and more important.
Correspondingly, the threshold peak decreases faster and/or merges with
the next peak at higher $\omega$ such that sometimes no peak remains at
the threshold. Also, the $\omega$-range where relevant continuum
contributions appear increases with decreasing temperature. All this is
demonstrated in Figs.\ \ref{fig:ao095l}, \ref{fig:ao097l} and  
\ref{fig:ao12l}, where we show our results for $J=0.95,\,0.97,\,1.0$ 
and $1.2\,$ with different $\omega$-ranges from 0 to 1-2. In the plots
for $J=1.0$ and 1.2 we have limited the peaks at $H=0.0001$ to enlarge
the other contributions for better visibility. The parameters for MEM
and the default model were $\Gamma=0.01$ and $\Delta\omega=0.001$. Since
$\omega_m$ sometimes extended up to 6 we have also tried
$\Delta\omega=0.0015$ or 0.002 and found no difference in the results.

%-------------------------------------------------------------------------
\setlength{\unitlength}{1cm}
\begin{picture}(10,9.5)
\put(1,0){
   \epsfig{bbllx=498,bblly=163,bburx=146,bbury=657,
       file=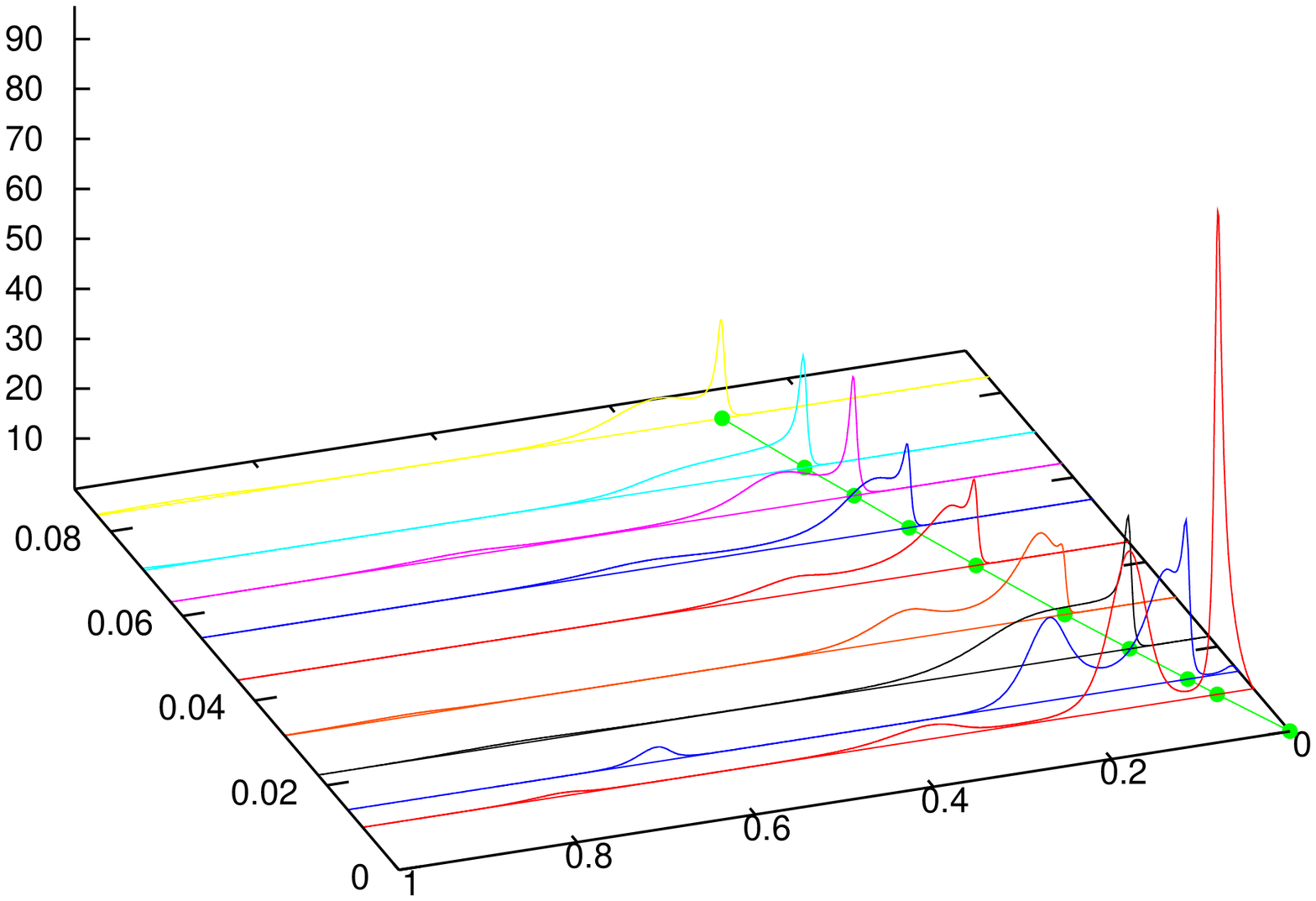,height=125mm,angle=-90}
          }
\put(1.8,8.5){\large{$A_L(\omega)$}}
\put(5.0,7.0){\large{$J=0.95$}}
\put(9.5,0){\large{$\omega$}}
\put(1,0.9){\large{$H^{1/2}$}}
\end{picture}
%------------------------------------------------------------------------
\vspace{3mm}
%------------------------------------------------------------------------
\begin{figure}[h]
\caption{The spectral function $A_L(\omega)$ at $J=0.95$ and fixed values 
of $H$ plotted versus $H^{1/2}$. For comparison the values of $2m_T$
(filled circles) are also shown.}
\label{fig:ao095l}
\end{figure}
%------------------------------------------------------------------------
\newpage
%----------------------------------------------------------------------------
\setlength{\unitlength}{1cm}
\begin{picture}(10,20)
\put(1,0){ 
   \epsfig{bbllx=498,bblly=163,bburx=146,bbury=657,
       file=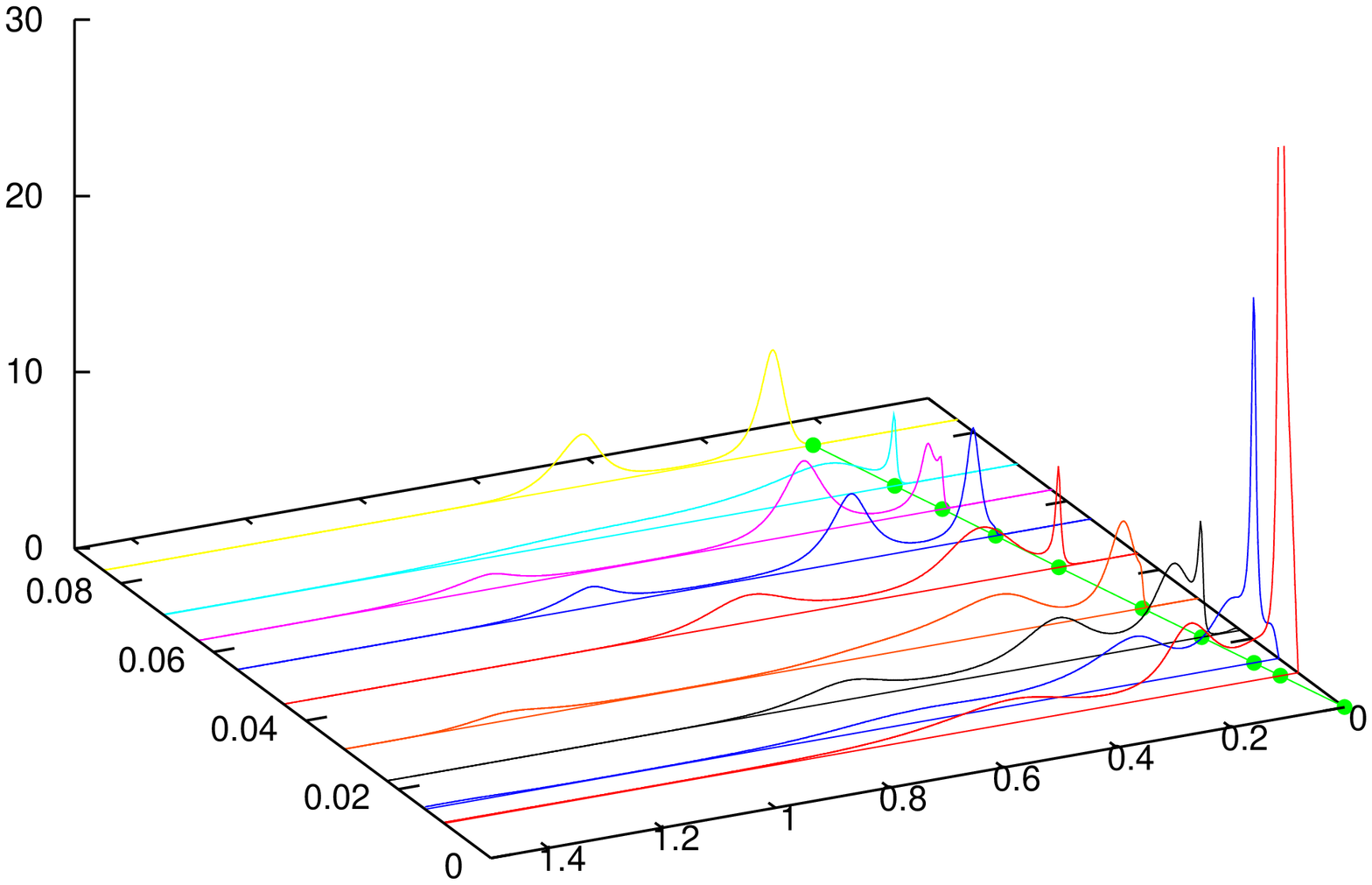,height=125mm,angle=-90}
          }
\put(1.8,8.5){\large{$A_L(\omega)$}}
\put(5.0,7.0){\large{$J=1.0$}}
\put(9.5,0){\large{$\omega$}}
\put(1,0.9){\large{$H^{1/2}$}}
\put(1,10){ 
   \epsfig{bbllx=498,bblly=163,bburx=146,bbury=657,
       file=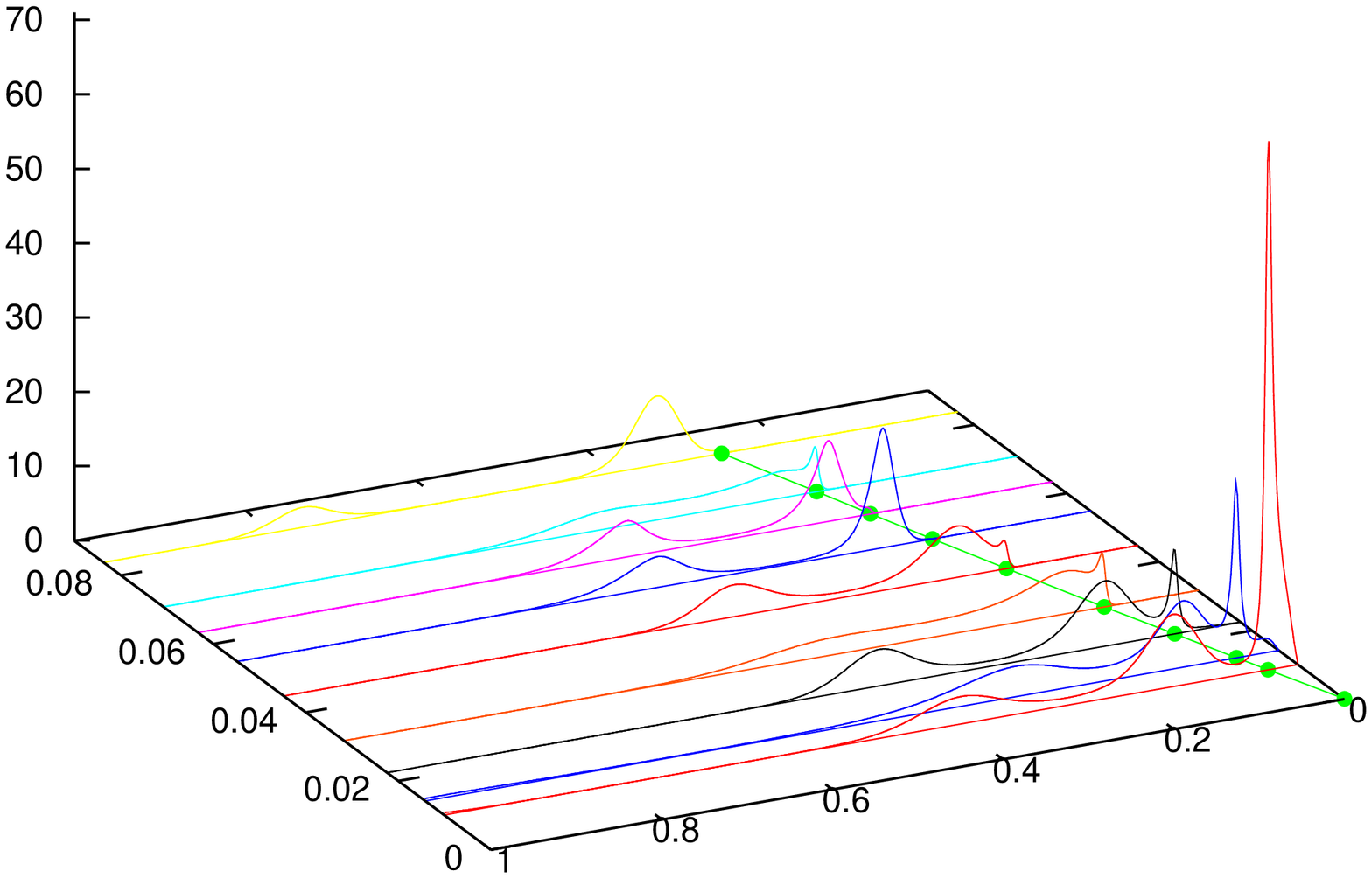,height=125mm,angle=-90}
          }
\put(1.8,18.5){\large{$A_L(\omega)$}}
\put(5.0,17.0){\large{$J=0.97$}}
\put(9.5,10){\large{$\omega$}}
\put(1,10.9){\large{$H^{1/2}$}}
\end{picture}
%------------------------------------------------------------------------
\vspace{3mm}
%------------------------------------------------------------------------
\begin{figure}[h]
\caption{The spectral function $A_L(\omega)$ at $J=0.97$ and 1.0 for
fixed values of $H$ plotted versus $H^{1/2}$. For comparison the values
of $2m_T$ (filled circles) are also shown.}
\label{fig:ao097l}
\end{figure}
\newpage
%----------------------------------------------------------------------------
\setlength{\unitlength}{1cm}
\begin{picture}(10,9.5)
\put(1,0){ 
   \epsfig{bbllx=498,bblly=163,bburx=146,bbury=657,
       file=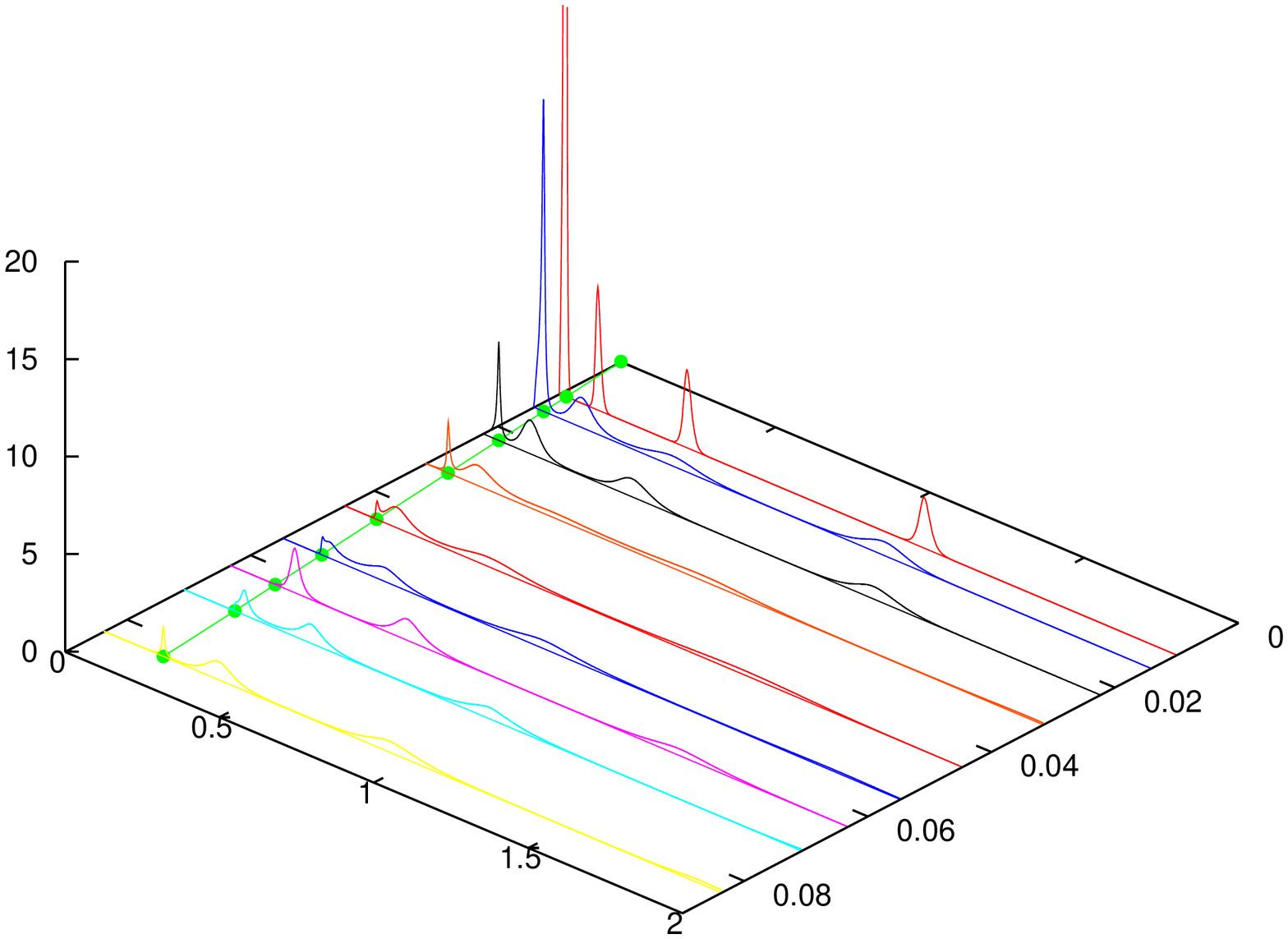,height=125mm,angle=-90}
          }
\put(1.4,7.2){\large{$A_L(\omega)$}}
\put(9.5,6.0){\large{$J=1.2$}}
\put(3,0.7){\large{$\omega$}}
\put(12,0.6){\large{$H^{1/2}$}}
\end{picture}
%------------------------------------------------------------------------
\vspace{3mm}
%------------------------------------------------------------------------
%------------------------------------------------------------------------
%%%%%%%%%%%%%%%%%%%%%%%%%%%%%%%%%%%%%%%%%%%%%%%%%%%%%%%%%%%%%%%%%%%%%%%%%

\section{Additional results}
\label{section:addresult}

%%%%%%%%%%%%%%%%%%%%%%%%%%%%%%%%%%%%%%%%%%%%%%%%%%%%%%%%%%%%%%%%%%%%%%%%%
\subsection{Test of the PP-relation}
\label{section:comparison}
\begin{figure}[t]
\begin{center}
   \epsfig{bbllx=36,bblly=135,bburx=620,bbury=553,
       file=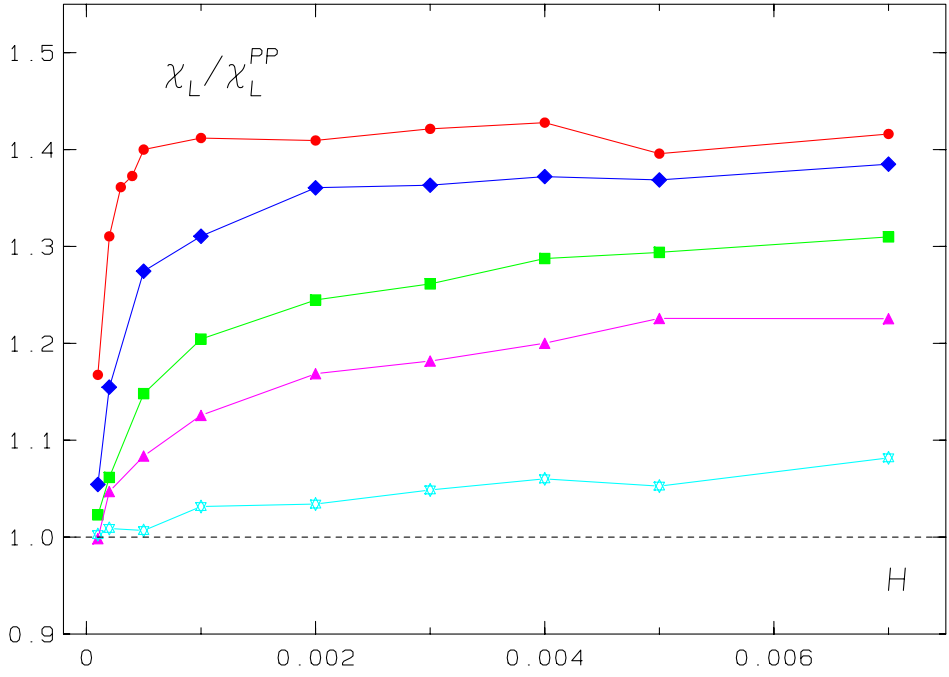, width=117mm}
\end{center}
\caption{The ratio $\chi_L/\chi_L^{PP}$ of the measured susceptibility 
$\chi_L$ to the result from the PP-relation, Eq.\ (\ref{cpp3}), 
for $J=0.94$ (circles), 0.95 (diamonds), 0.97 (squares), 1.0 (triangles)
and 1.2 (davidstars) versus $H$.}
\label{fig:cppc}
\end{figure}

\begin{figure}[ht]
\begin{center}
   \epsfig{bbllx=36,bblly=139,bburx=621,bbury=557,
       file=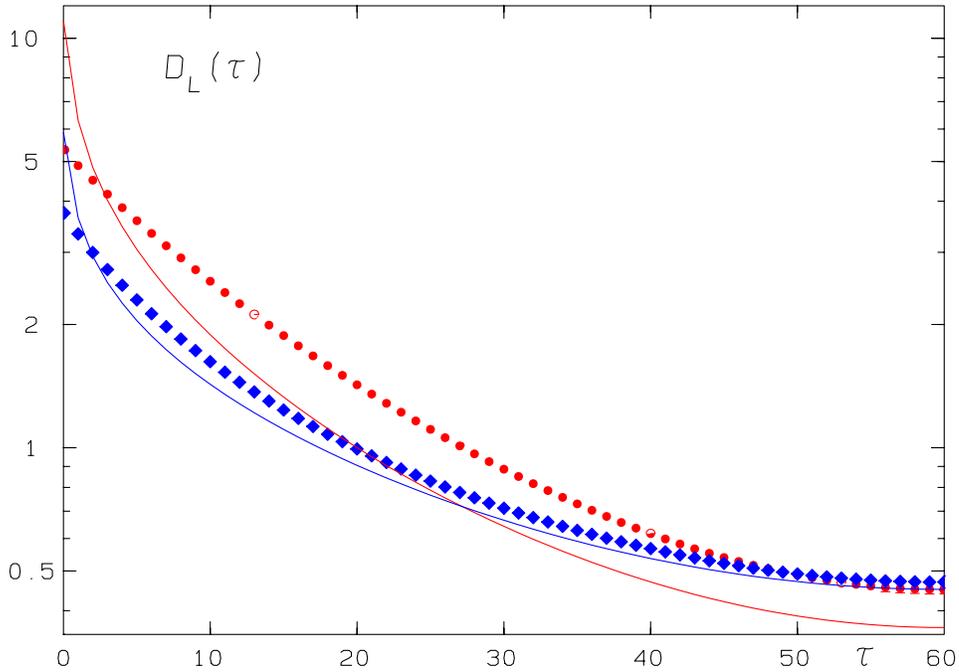, width=117mm}
\end{center}
\caption{The correlator $D_L(\tau)$ for $J=0.94$ (circles)
and 0.95 (diamonds) at $H=0.0001$. The lines show the respective results
from the PP-relation (\ref{dpp2}).}
\label{fig:mla0001}
\end{figure}

\begin{figure}[t]
\begin{center}
   \epsfig{bbllx=36,bblly=136,bburx=622,bbury=555,
       file=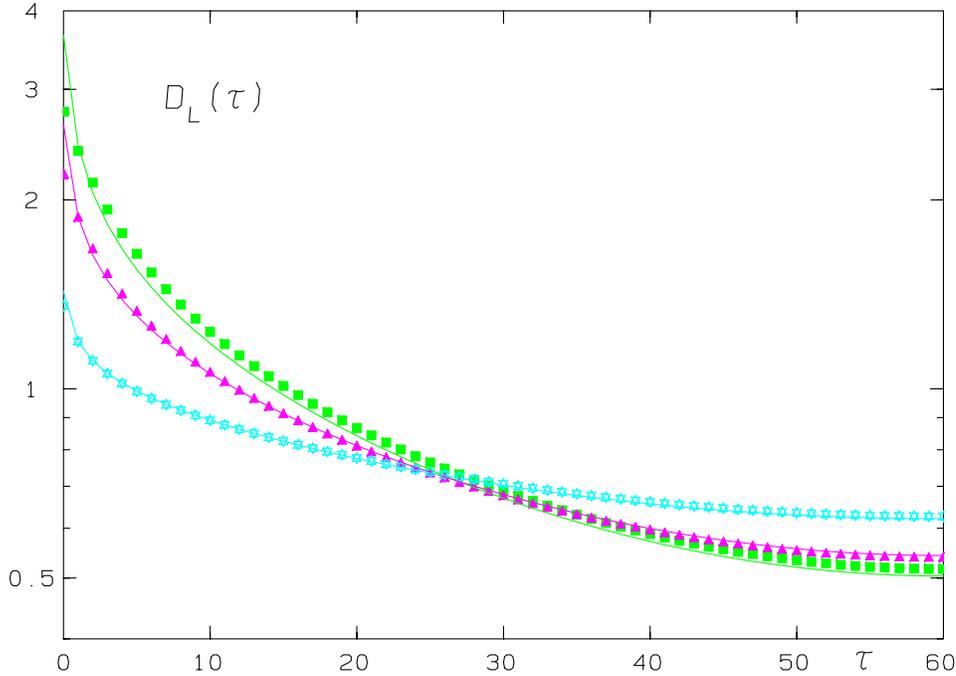, width=117mm}
\end{center}
\caption{The correlator $D_L(\tau)$ for $J=0.97$ (squares), 1.0 
(triangles) and 1.2 (davidstars) at $H=0.0001$. The lines show the 
respective results from the PP-relation (\ref{dpp2}).}
\label{fig:mlb0001}
\end{figure}

%%%%%%%%%%%%%%%%%%%%%%%%%%%%%%%%%%%%%%%%%%%%%%%%%%%%%%%%%%%%%%%%%%%%%%%%%
\n In Section \ref{section:transspec} we have shown that the transverse 
correlators have the Gaussian form and we have determined the respective
transverse masses $m_T$. This enables us now to test the relation of
Patashinskii and Pokrovskii which we have discussed in Section 
\ref{section:tlrel}\,. In Fig.\ \ref{fig:cppc} we plot the ratio of 
the measured longitudinal susceptibility $\chi_L$ to the result 
$\chi_L^{PP}$ of Eq.\ (\ref{cpp3}) as a function of $H$ for all our
couplings $J>J_c$ in the low temperature phase. As expected, the ratio
is close to 1 for very small $H$-values. For all couplings the ratio
increases with increasing external field and reaches essentially a
constant value. With decreasing temperature (increasing $J$) the ratio
approaches 1 from above and at our lowest temperature $T=1/1.2$ we are
for all $H$-values already below 1.1\,. The test of $\chi_L^{PP}$ is
of course a global one. More details are obtained from a direct 
comparison of the longitudinal correlator $D_L(\tau)$ to its counterpart
from Eq.\ (\ref{dpp2}). In Fig.\ \ref{fig:mla0001} and \ref{fig:mlb0001}
we show all our measured correlation functions for $H=0.0001$ in the low 
temperature phase together with
%------------------------------------------------------------------------
%------------------------------------------------------------------------
%------------------------------------------------------------------------
%------------------------------------------------------------------------
\newpage
%------------------------------------------------------------------------
%------------------------------------------------------------------------
\n the respective results from (\ref{dpp2}). In Fig.\ \ref{fig:mla0001} 
we plot the correlators for $J=0.94$ and 0.95, the two couplings
closest to the critical point. Whereas for $J=0.94$ the prediction from
Eq.\ (\ref{dpp2}) is still definitely different from the data, the line
and the data are close to each other already for $J=0.95$, apart from the
small $\tau$-region. In Fig.\ \ref{fig:mlb0001} we observe with decreasing
temperature a further approach, at $J=1.2$ the prediction coincides fully 
with the data. In the high temperature region the relation (\ref{PP})
obviously cannot hold for $H\to 0$\,: there is no spontaneous magnetization
and the longitudinal and transverse correlation functions must agree
with each other. Accordingly, $\chi_L$ and $\chi_L^{PP}$ can differ by a 
factor of $10^2$ for small external fields.
%%%%%%%%%%%%%%%%%%%%%%%%%%%%%%%%%%%%%%%%%%%%%%%%%%%%%%%%%%%%%%%%%%%%%%%%%
\subsection{The stiffness}
\label{section:stiff}
\begin{figure}[t]
\begin{center}
   \epsfig{bbllx=36,bblly=133,bburx=621,bbury=552,
       file=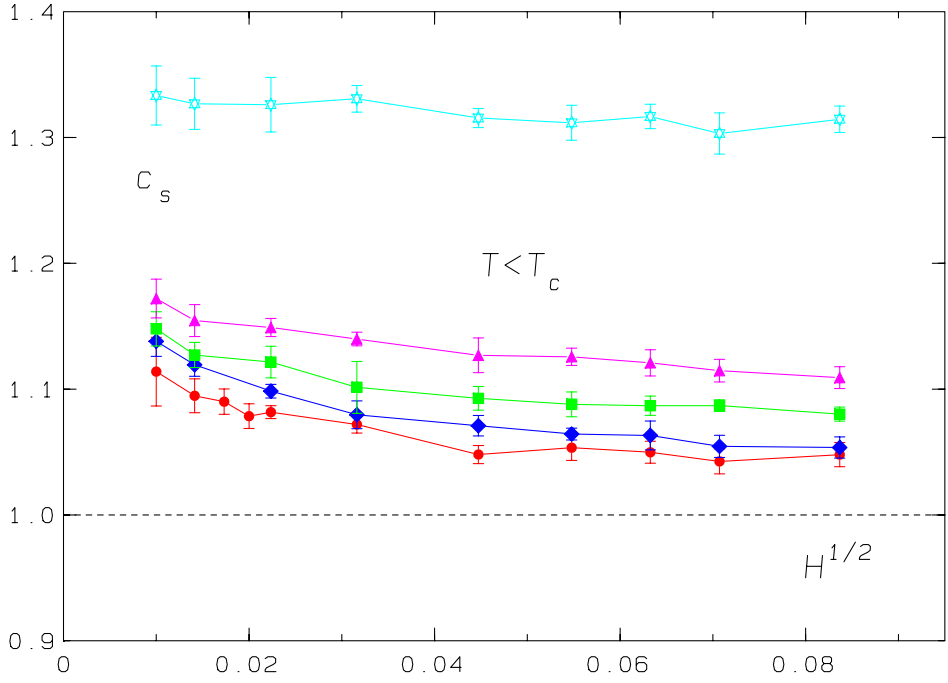, width=117mm}
\end{center}
\caption{The stiffness $c_s$ versus $H^{1/2}$ in the low temperature
phase for $J=0.94$ (circles), 0.95 (diamonds), 0.97 (squares), 
1.0 (triangles) and 1.2 (davidstars). The lines are intended to guide
the eye.}
\label{fig:cscold}
\end{figure}

\begin{figure}[b]
\begin{center}
   \epsfig{bbllx=40,bblly=179,bburx=630,bbury=514,
       file=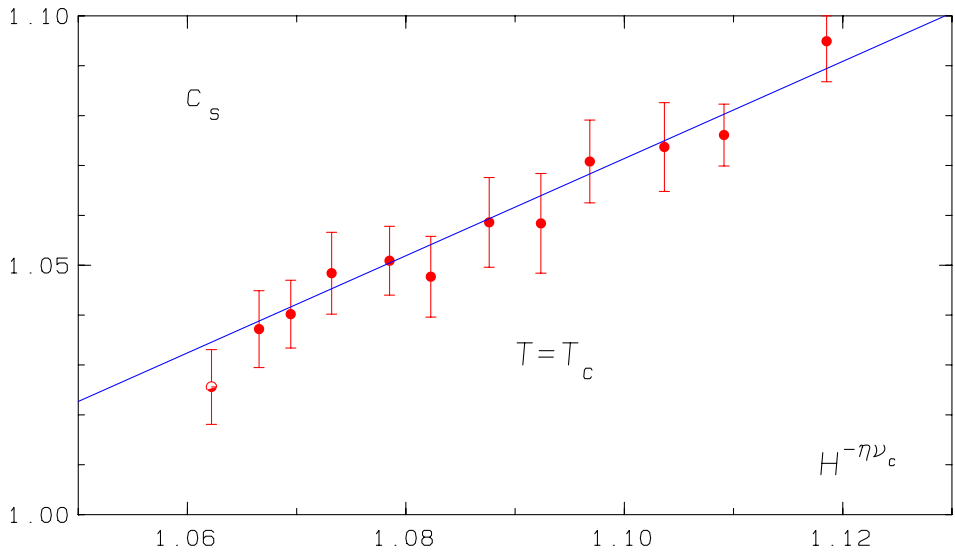, width=117mm}
\end{center}
\caption{The stiffness $c_s$ on the critical line versus $H^{-\eta\nu_c}$
and the prediction (line) from Eq.\ (\ref{csattc}). }
\label{fig:csjc}
\end{figure}

\begin{figure}[t]
\begin{center}
   \epsfig{bbllx=46,bblly=181,bburx=631,bbury=515,
       file=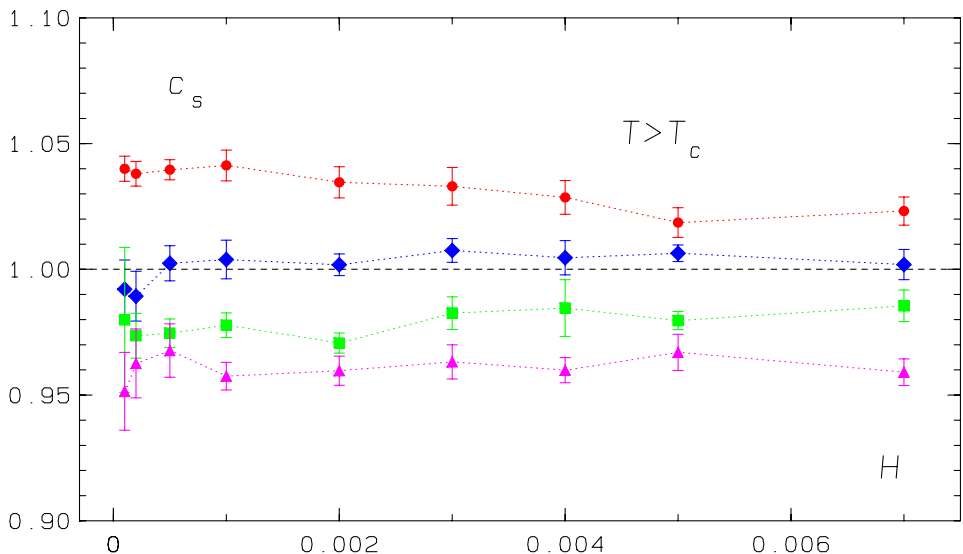, width=117mm}
\end{center}
\caption{The stiffness $c_s$ as a function of $H$ in the high temperature
phase for $J=0.93$ (circles), 0.92 (diamonds), 0.91 (squares) 
and 0.90 (triangles). The dotted lines are drawn to guide the eye.}
\label{fig:cshot}
\end{figure}

%%%%%%%%%%%%%%%%%%%%%%%%%%%%%%%%%%%%%%%%%%%%%%%%%%%%%%%%%%%%%%%%%%%%%%%%%
\n In Sections \ref{section:tlrel} and \ref{section:Criti} we have
discussed the behaviour of the stiffness $c_s$. From our results for
the transverse mass $m_T$ and the transverse susceptibility $\chi_T$ we
can calculate $c_s$ using Eq.\ (\ref{mt})\,. In Fig.\ \ref{fig:cscold}
we show the stiffness as a function of $H^{1/2}$ (see Eq.\ (\ref{csasym})
\,) for fixed temperatures below $T_c$. As expected, the stiffness 
increases with decreasing temperature. With decreasing external field
the stiffness grows slightly, however for $H\to 0$ a finite value will
be reached. This is different on the critical line, where 
\be
c_s = H_0\,[g_{\xi}^T(0)]^2 h^{-\eta\nu_c} =\, 0.9740\, H^{-\eta\nu_c}~,
\label{csattc}
\ee
and $c_s$ is weakly diverging for $H\to 0$. We have calculated the
prefactor of $H^{-\eta\nu_c}$ from the critical parameters given in 
Ref.\ \cite{Engels:2003nq}. The prediction from the last equation is
confirmed by our data as can be seen in Fig.\ \ref{fig:csjc}. In the 
high temperature region  
%------------------------------------------------------------------------
%------------------------------------------------------------------------
%------------------------------------------------------------------------
%------------------------------------------------------------------------
\newpage
\n $c_s$ is close to one and only slowly varying
with the external field. This is evident from Fig.\ \ref{fig:cshot}, where
we have plotted $c_s$ as a function of $H$ and not $H^2$ to avoid a too
dense plot at small $H$.
%------------------------------------------------------------------------
%------------------------------------------------------------------------
%%%%%%%%%%%%%%%%%%%%%%%%%%%%%%%%%%%%%%%%%%%%%%%%%%%%%%%%%%%%%%%%%%%%%%%%%
\subsection{Scaling of the longitudinal correlation length}
\label{section:gxil}
\begin{figure}[t]
\begin{center}
   \epsfig{bbllx=36,bblly=133,bburx=621,bbury=552,
       file=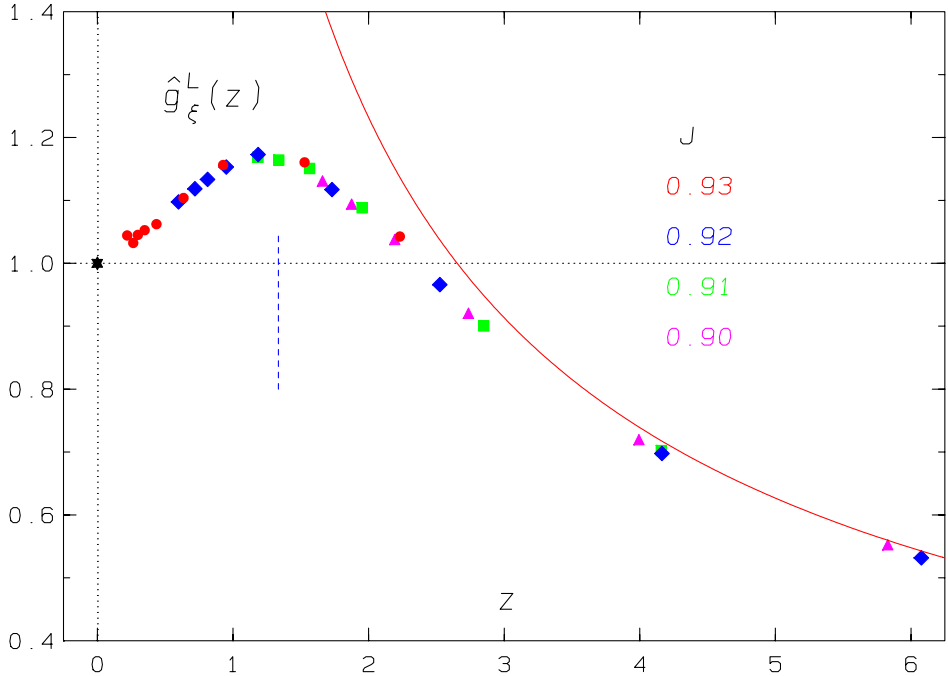, width=117mm}
\end{center}
\caption{The normalized scaling function $\hat g_{\xi}^L(z)$ in the high
temperature phase for $J=0.93$ (circles), 0.92 (diamonds), 0.91 (squares) 
and 0.90 (triangles). The dashed line indicates the peak position of
$\chi_L$, the solid line is the asymptotic form (\ref{gxiasy}).}
\label{fig:glscale}
\end{figure}

%%%%%%%%%%%%%%%%%%%%%%%%%%%%%%%%%%%%%%%%%%%%%%%%%%%%%%%%%%%%%%%%%%%%%%%%%
\n In Ref.\ \cite{Engels:2003nq} the scaling function of the longitudinal
correlation length was calculated from the directly measured values of
$\xi_L$. For $T\ge T_c$ the longitudinal mass $m_L$ is the inverse of the
longitudinal correlation length. Such an identification is not possible
in the low temperature range where we have a whole continuum of 
contributing $m_L$ or $\omega$-values. In the high temperature region
however we can use the relation
\be
\hat g_{\xi}^L(z) = \frac{h^{\nu_c}}{g^L_{\xi}(0)m_L}~,
\label{gxilh}
\ee
to calculate the scaling function from our $m_L$-data. The result is 
shown in Fig.\ \ref{fig:glscale} together with the asymptotic form
for large $z$-values from Eq.\ (\ref{gxiasy}). Our data are scaling
surprisingly well, better than the former results in Fig.\ 9 of Ref.\ 
\cite{Engels:2003nq}, in particular in the small $z$-region. One reason
for the improvement may be the increase in the lattice size used, the
other one is certainly the safer determination of $m_L$ as compared to
that of $\xi_L$. As in Ref.\ \cite{Engels:2003nq} we note the similarity
in the form of this scaling function and that of the longitudinal
susceptibility. Consequently, both functions peak at about the same 
value of $z$, $z_p=1.335\,$. 
\clearpage
%------------------------------------------------------------------------
%------------------------------------------------------------------------
%%%%%%%%%%%%%%%%%%%%%%%%%%%%%%%%%%%%%%%%%%%%%%%%%%%%%%%%%%%%%%%%%%%%%%%%%
\subsection{Magnetization and susceptibilities at $T_c$}
\label{section:delta}
\begin{figure}[t]
\begin{center}
   \epsfig{bbllx=36,bblly=183,bburx=621,bbury=518,
       file=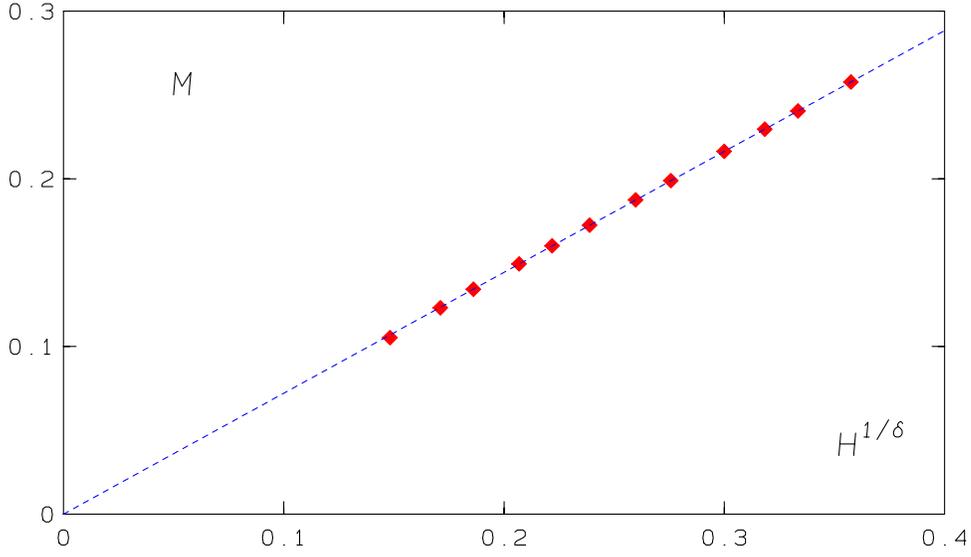, width=117mm}
\end{center}
\caption{The magnetization $M$ (diamonds) as a function of $H^{1/\delta}$ 
on the critical line. The dashed line is given by Eq.\ (\ref{mcrh})
with the parameters from Ref.\ \cite{Engels:2003nq}.}
\label{fig:Matjc}
\end{figure}

\begin{figure}[ht]
\begin{center}
   \epsfig{bbllx=36,bblly=177,bburx=621,bbury=511,
       file=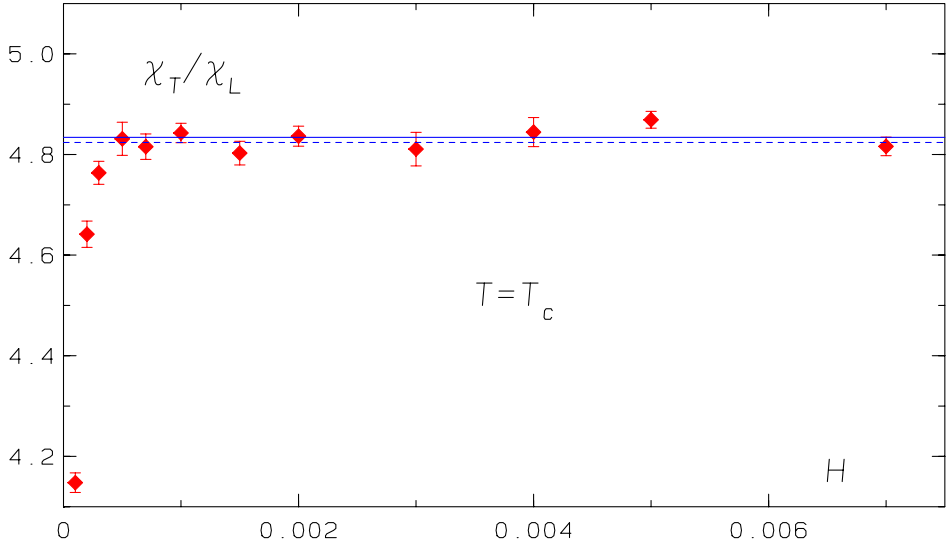, width=117mm}
\end{center}
\caption{The ratio $\chi_T/\chi_L$ (diamonds) versus $H$ 
on the critical line. The dashed line is $\delta=4.824$
from Ref.\ \cite{Engels:2003nq}, the solid line the fit result
$\delta=4.834$.}
\label{fig:chiatjc}
\end{figure}

%%%%%%%%%%%%%%%%%%%%%%%%%%%%%%%%%%%%%%%%%%%%%%%%%%%%%%%%%%%%%%%%%%%%%%%%%
\n The behaviour of the magnetization $M$ and the susceptibilities
$\chi_L$ and $\chi_T$ on the critical line $T=T_c$ is controlled by
the critical exponent $\delta$ or rather its inverse as is evident from
Eqs.\ (\ref{mcrh}) and (\ref{chicrh}). In Fig.\ \ref{fig:Matjc} we show
our data for $M$ at $T_c$ as a function of $H^{1/\delta}$, where 
$\delta=4.824$, the value of \cite{Engels:2003nq} was used. We compare
the data with the result from Eq.\ (\ref{mcrh}) with $B^c=0.721$, again 
from \cite{Engels:2003nq}. Except for the point at $H=0.0001$ which
perhaps shows some finite size effect, all data agree very well with the
line. A straight line fit to the data for $\ln M$ as a function of 
$\ln H$ without the first point leads to a somewhat more accurate result
\be
B^c = 0.7198\pm 0.0005~,\quad {\rm and}\quad \delta = 4.831\pm 0.003~.
\label{resM}
\ee 
The ratio $\chi_T/\chi_L$ offers a straightforward way to determine
$\delta$\,: on the critical line it should coincide with the exponent.
The drawback is however that the ratio has larger errors and that it
shows definite finite size effects at small $H$-values. The latter are
due to the longitudinal susceptibility. In Fig.\ \ref{fig:chiatjc} we
show our data for the ratio. The values for $H\le 0.003$ are still 
affected by the volume dependence. In order to determine $\delta$ we
have therefore used only the points from $H\ge 0.0005$. We find
\be
\delta= 4.834\pm 0.007~.
\label{resrat}
\ee
%------------------------------------------------------------------------
%------------------------------------------------------------------------
%------------------------------------------------------------------------
%------------------------------------------------------------------------
%%%%%%%%%%%%%%%%%%%%%%%%%%%%%%%%%%%%%%%%%%%%%%%%%%%%%%%%%%%%%%%%%%%%%%%%%

\section{Summary and conclusions}
\label{section:sumcon}

%%%%%%%%%%%%%%%%%%%%%%%%%%%%%%%%%%%%%%%%%%%%%%%%%%%%%%%%%%%%%%%%%%%%%%%%%
\n In our paper we have investigated the transverse and longitudinal
2-plane spin correlation functions of the three-dimensional O(4)-invariant 
non-linear $\sigma$-model. The data were obtained from simulations on a
cubic lattice with linear extension $L=120$ for small external fields $H$
and temperatures below, above and at the critical temperature $T_c\,$.
The main objective was the determination of the spectral functions of
the correlators. In particular we were interested in the identification
of the lowest states of the spectra $m_{T,L}$, which relate to the
correlation lengthes by $\xi_{T,L}=1/m_{T,L}$, the influence of possible 
higher states on the correlation functions, and the interplay of the
transverse and longitudinal channels. In order to calculate the spectral 
functions we used Bryan's algorithm for the maximum entropy method with
a modified kernel which improves numerical stability at small $\omega$.

\n As a result we find in the transverse case a {\em single} sharp peak
in the spectral function for each of our $T$ and $H$-values. The
correlator is that of a free particle with mass $m_T$, that is it has 
the known Gaussian form. We confirm therefore also corresponding 
assumptions of spin-wave theory about the dominance of long-wavelegth
transverse fluctuations (i.e. small $m_T$) of Gaussian type for small
fields below $T_c$\,. In the very high temperature region we observe a 
single sharp peak as well in the longitudinal spectrum. On approaching
the critical point from above the peak broadens somewhat and at $T_c$
its position $m_L$ coincides with $2m_T$ for all our $H-$values.
In the low temperature region $T<T_c$ we still find a significant peak 
at $\omega=2m_T$. At higher $\omega-$values a continuum of states 
with several peaks of decreasing heights appears. This is expected from
a relation of Patashinskii and Pokrovskii between the longitudinal and 
the transverse correlation functions. However, a comparison of the
relation with the correlation function data reveals that there must be 
additional contributions to the longitudinal spectra at higher external
fields. With decreasing temperatures these additive contributions 
gradually disappear.  
 
\n The influence of the longitudinal spin component on the transverse ones
is characterized by the stiffness $c_s$\,. As expected we find an increase
of $c_s$ with decreasing temperature below $T_c$\,. Above the critical
temperature the stiffness is close to one and only weakly dependent on
the external field. At $T_c$ we verify the predicted critical behaviour
$c_s \sim H^{-\eta\nu_c}$. With our results for $m_L$ for the high 
temperature region we have calculated the scaling function of the
longitudinal correlation length. The new data scale surprisingly well, 
better than those of Ref.\ \cite{Engels:2003nq}, which were derived 
from direct measurements of the correlation length. As a last check we
have compared the data for the magnetization and the susceptibilities
to the predicted critical behaviour at $T_c$. We find again agreement 
with the former values $B^c=0.721(2)$ and $\delta=4.824(9)$ from 
\cite{Engels:2003nq}. A new fit to our data for the magnetization leads
however to the more accurate numbers $B^c=0.7198(5)$ and $\delta=4.831(3)$. 
%%%%%%%%%%%%%%%%%%%%%%%%%%%%%%%%%%%%%%%%%%%%%%%%%%%%%%%%%%%%%%%%%%%%%%%%%%
\vskip 0.2truecm
\noindent{\Large{\bf Acknowledgments}}

%%%%%%%%%%%%%%%%%%%%%%%%%%%%%%%%%%%%%%%%%%%%%%%%%%%%%%%%%%%%%%%%%%%%%%%%%%
\n We thank Frithjof Karsch for his suggestion to start this work and
his constant interest. We are indebted to Olaf Kaczmarek for numerous 
stimulating discussions on the maximum entropy method and to Ines
Wetzorke for the permission to use her MEM-program.

%%%%%%%%%%%%%%%%%%%%%%%%%%%%%%%%%%%%%%%%%%%%%%%%%%%%%%%%%%%%%%%%%%%%%%%%%%
%%%%%%%%%%%%%%%%%%%%%%%%%%%%%%%%%%%%%%%%%%%%%%%%%%%%%%%%%%%%%%%%%%%%%%%%%%

%\clearpage
%%%%%%%%%%%%%%%%%%%%%%%%%%%%%%%%%%%%%%%%%%%%%%%%%%%%%%%%%%%%%%%%%%%%%%%%%%%%%%%%
\end{document}